\documentclass[11pt,letterpaper]{article}
\usepackage[margin=1in]{geometry}

\usepackage[T1]{fontenc}
\usepackage{lmodern}
\usepackage{amsmath,amssymb,amsthm,mathtools}
\usepackage{algorithm,algpseudocode}
\usepackage{booktabs,array}
\usepackage[section]{placeins}
\usepackage{microtype}
\usepackage[dvipsnames]{xcolor}
\usepackage{tikz}
\usepackage{booktabs, array, tabularx, colortbl}
\usepackage{comment}
\usepackage{mathrsfs}
\usetikzlibrary{shapes.geometric}

\usetikzlibrary{arrows.meta,positioning,calc,decorations.pathreplacing}

\usepackage[backend=biber,style=alphabetic,sorting=nyt,maxbibnames=99,giveninits=false,isbn=false]{biblatex}
\usepackage[colorlinks=true,linkcolor=blue!45!black,citecolor=blue!45!black,urlcolor=blue!45!black]{hyperref}

\newcommand{\nb}[3]{{\colorbox{#2}{\bfseries\sffamily\scriptsize\textcolor{white}{#1}}}{\textcolor{#2}{\sf\small\textit{#3}}}}

\newcommand{\mat}[1]{\nb{mat}{blue}{#1}}
\newcommand{\anna}[1]{\nb{anna}{ForestGreen}{#1}}

\newtheorem{theorem}{Theorem}[section]
\newtheorem{lemma}[theorem]{Lemma}
\newtheorem{proposition}[theorem]{Proposition}
\newtheorem{corollary}[theorem]{Corollary}

\newtheorem{requirement}[theorem]{Requirement}
\newtheorem{assumption}[theorem]{Assumption}
\theoremstyle{definition}
\newtheorem{definition}[theorem]{Definition}
\newcommand{\E}{\mathbb E}

\newcommand{\ind}{\mathbb I}

\newcommand{\Gtar}{\mathcal G_{\mathrm{tar}}}

\newcommand{\Gqry}{\mathcal G_{\mathrm{qry}}}

\newcommand{\OPT}{\mathrm{OPT}}

\newcommand{\Dyad}{\mathcal I^\star}

\newcommand{\gap}{\operatorname{gap}}

\usepackage{thm-restate}

\title{Optimal Regret for Profit Maximization in Bilateral Trade%\\ With One-Bit Feedback
}

\author{
Matteo Castiglioni\thanks{The authors are listed in alphabetical order.}\\
Politecnico di Milano\\
\and
Anna Lunghi\footnotemark[1]\\
Politecnico di Milano\\
\and
Alberto Marchesi\footnotemark[1]\\
Politecnico di Milano
}

\date{\texttt{name.surname@polimi.it}
}

\begin{document}

\maketitle

\begin{abstract}\noindent
Learning to profit from trade requires learning both sides of a market: what a seller is willing to accept and what a buyer is willing to pay. We study repeated bilateral trade, where an intermediary posts a purchase price to a seller and a sale price to a buyer whose valuations are drawn from an unknown joint distribution. Trade occurs only if both parties accept their respective prices, in which case the intermediary earns the difference between the two prices. The intermediary observes only whether trade occurs, without observing either valuation.

We show that learning both sides of the market is no harder, in terms of regret, than learning just one. Specifically, we design an algorithm achieving regret $\widetilde{\mathcal{O}}(T^{2/3})$ against the best fixed pair of prices, matching, up to logarithmic factors, the classical lower bound for one-dimensional dynamic pricing. Our result shows that the classical one-dimensional regret rate survives the transition to two dimensions, raising the question of whether this absence of a curse of dimensionality extends to arbitrarily high dimensions.

\end{abstract}

\newpage
\tableofcontents
\newpage

\section{Introduction}\label{sec:introduction}

Bilateral trade is a fundamental model of economic exchange under private information. A seller holds an item that a buyer may wish to purchase, and an intermediary facilitates the transaction by offering a purchase price to the seller and a sale price to the buyer. Classical mechanism design studies how private valuations constrain the efficiency of such exchanges~\cite{MyersonS83}. More recently, starting with the seminal work of~\cite{cesa2021regret}, considerable attention has been devoted to learning bilateral-trade mechanisms when the distribution of agents' valuations is unknown \cite{cesa2023repeated,azar2022alpha,bernasconi2024no,lunghi2025better,chen2025tight}.

In this paper, we study profit maximization in repeated bilateral trade over $T$ rounds. At each round $t$, a seller holds an item with private valuation $S_t\in[0,1]$, and a buyer values it at $B_t\in[0,1]$. The valuation pairs $(S_t,B_t)$ are drawn independently from an unknown joint distribution. A learner, acting as an intermediary, posts a purchase price $p_t\in[0,1]$ to the seller and a sale price $q_t\in[0,1]$ to the buyer, without observing their valuations. The seller accepts if $S_t\le p_t$, and the buyer accepts if $B_t\ge q_t$. If both accept, the intermediary buys the item for $p_t$ and sells it for $q_t$, earning a profit of $q_t-p_t$. Otherwise, no trade occurs and no payment is made.

After posting its prices, the learner observes only whether trade occurs, namely,
\[
Y_t=\ind\{S_t\le p_t,\ B_t\ge q_t\},
\]
which also determines its realized profit $(q_t-p_t)Y_t$. The learner's goal is to minimize the regret with respect to the profit of the optimal pair of prices. 
%The learner's objective is to maximize its expected 

To highlight the connection with learning cumulative distribution functions (CDFs), we introduce the transformed valuations $X_t=(S_t,1-B_t)$ and actions $x_t=(p_t,1-q_t)$. Under this representation, trade occurs exactly when $X_t\le x_t$, where inequalities are understood coordinatewise. Let $F(x)=\mathbb P(X_t\le x)$ denote the cumulative distribution function (CDF) of the transformed valuations. The learner then observes a Bernoulli random variable with mean $F(x_t)$, and the expected profit of an action $x=(x_1,x_2)$ is
\[
r(x)=(1-x_1-x_2)F(x).
\]
Thus, our problem can be viewed as learning to maximize the product of a known affine function and an unknown bivariate CDF, using only one-bit observations of the latter.

Despite its connection with this broader class of problems, the main focus of this paper is bilateral trade. Most of the literature on learning in bilateral trade has focused on maximizing the Gain From Trade (GFT), which measures the economic surplus generated by a transaction, rather than the intermediary's profit. A central challenge in this setting is the absence of bandit feedback, which has motivated a long line of research aimed at overcoming this limitation (see Section \ref{sec:related}).

Profit maximization, by contrast, has received considerably less
attention, despite its natural economic interpretation.
Recent work~\cite{di2025nearly,di2026profit} has investigated
this objective in more general classes of trading mechanisms,
rather than the fixed-price setting considered here.
One possible explanation for this limited attention is that
profit maximization naturally admits bandit feedback: the intermediary can compute its realized profit from the posted prices and the observation of whether trade occurs. The availability of bandit feedback makes the problem appear to be a standard instance of one-sided Lipschitz bandits, for which classical techniques readily apply. Indeed, a straightforward discretization of the two-dimensional price space yields an algorithm with regret $\widetilde{\mathcal{O}}(T^{3/4})$. To see this, consider a grid with mesh size $\varepsilon$, containing $\mathcal{O}(\varepsilon^{-2})$ pairs of prices. Exploiting the one-sided Lipschitz property of the expected profit function, discretization incurs at most $\mathcal{O}(\varepsilon T)$ regret. Running a standard multi-armed bandit algorithm over the grid therefore yields regret $\widetilde{\mathcal{O}}(\varepsilon T+\sqrt{T/\varepsilon^2})$, which is minimized by choosing $\varepsilon=T^{-1/4}$.

A recent result by~\cite{castiglioni2026sample} suggests that this standard approach may be suboptimal and that better regret rates might be achievable. Specifically, the authors show that the sample complexity of uniformly learning CDFs is essentially independent of the number of variables, for any constant number of dimensions. As a corollary, learning a near-optimal pair of prices in bilateral trade requires no more samples than learning a near-optimal price in classical dynamic pricing. 

Whether this sample efficiency translates into improved regret guarantees, however, is far from immediate. Indeed, the sample-complexity guarantees of~\cite{castiglioni2026sample} yield regret $\widetilde{\mathcal{O}}(T^{3/4})$ through an explore-then-commit strategy, matching the rate obtained through standard Lipschitz bandit techniques. Improving on this rate requires not only sample-efficient learning, but also ensuring that the actions played during learning are sufficiently close to optimal. Exploiting the structure of bivariate CDFs,~\cite{CLM26b} recently made progress in this direction, breaking the $T^{3/4}$ barrier and achieving regret $\widetilde{\mathcal{O}}(T^{7/10})$.

Despite this progress, a fundamental question remains: is learning two prices intrinsically harder than learning a single one? In classical single-buyer dynamic pricing, the optimal regret rate is $\widetilde{\Theta}(T^{2/3})$~\cite{kleinberg2003value}. Since this problem is a special case of bilateral trade, obtained by fixing the seller's valuation to zero, the same lower bound applies to our setting. The gap between the lower bound of $T^{2/3}$ and the best known upper bound of $T^{7/10}$ thus leaves open the following question:

\begin{quote}
\emph{Can profit maximization in bilateral trade achieve the same regret rate as classical  dynamic pricing?}
\end{quote}

We answer this question affirmatively. We design an algorithm achieving $\widetilde{\mathcal{O}}(T^{2/3})$ expected regret for arbitrary joint distributions of the seller's and buyer's valuations.

\begin{theorem}[Informal version of Theorem~\ref{thm:main}]
There exists an algorithm for profit maximization in the repeated bilateral trade problem with expected regret
\[
    R_T\leq \widetilde{\mathcal{O}}\bigl(T^{2/3}\bigr).
\]
\end{theorem}
Together with the classical lower bound for dynamic pricing~\cite{kleinberg2003value}, this establishes the optimal regret rate up to logarithmic factors.

Our result shows that moving from one price to two does not incur a curse of dimensionality in regret minimization: learning to profit from a two-sided market is no harder than learning to sell to a single buyer. More broadly, it provides further evidence that the structure of CDFs can be exploited to overcome the dimensional dependence typically encountered in multidimensional (one-sided) Lipschitz bandits.

While we settle this question for bilateral trade, whether the absence of a curse of dimensionality extends to higher dimensions remains open.  In particular, can the
$\widetilde{\mathcal{O}}(T^{2/3})$ regret rate be achieved
for multidimensional CDF-based objectives,
such as those arising in markets with an arbitrary constant
number of buyers and sellers? Or does the dimension-independent
sample complexity established by~\cite{castiglioni2026sample}
fail to translate into dimension-independent regret guarantees?

\subsection{Challenges and Techniques}
\label{sec:overview}

Our algorithm works on a uniform discretization of the
two-dimensional price space, with $K$ points per coordinate. This reduces the problem to learning the best among $K^2$ price pairs, at an additional
regret cost of $\mathcal{O}(T/K)$. However, treating these pairs as unrelated actions
would overlook the structure of the problem: their trade
probabilities are determined by the same underlying
bivariate CDF.

With this discretization in place, our starting point is the procedure of \cite{castiglioni2026sample}, which uniformly
estimates the underlying CDF, and hence the expected profit,
over the grid to additive precision $\varepsilon$ using
$\widetilde{\mathcal O}(\varepsilon^{-3})$ samples.
This sample complexity suggests a route to $\widetilde {\mathcal O}(T^{2/3})$ regret, but several technical obstacles stand along the way.

A natural approach to translating this sample-complexity
result into regret guarantees is to incorporate
the procedure into a phased successive-elimination framework. For a geometrically decreasing sequence of precisions $\varepsilon_i$, phase $i$ refines the profit estimates of the active price pairs to precision $\varepsilon_i$, then eliminates those that are sufficiently suboptimal. If each phase used $\widetilde {\mathcal O}(\varepsilon_i^{-3})$ samples and queried only price pairs with suboptimality gap $\mathcal O(\varepsilon_i)$, its regret would be $\widetilde {\mathcal O}(\varepsilon_i^{-2})$. Choosing the final precision $\varepsilon_m$ at scale $T^{-1/3}$, and then committing to the best estimated pair, would give total regret
\[
\sum_{i=1}^{m}\widetilde {\mathcal O}(\varepsilon_i^{-2})
+\mathcal O(T\varepsilon_m)
=\widetilde {\mathcal O}(T^{2/3}).
\]

The main obstacle is that this argument requires more than uniform learning over a complete grid: it requires uniformly estimating profits over an arbitrary target set of points while querying only points in that set.
Eliminating actions can destroy the grid structure that makes the original procedure efficient, making it challenging to retain the same sample complexity under these restrictions.

Rather than establishing such a uniform-learning guarantee
for arbitrary active sets, we circumvent this obstacle by
modifying the successive-elimination framework to distinguish
between the points whose profits must be estimated and those
available for sampling.
Specifically, in each phase, our algorithm maintains a target set of promising price pairs whose profit estimates must be refined, together with a larger queryable set of price pairs whose suboptimality gaps remain small. The key idea is to retain some points already known to be suboptimal as auxiliary query points. Their presence provides the geometric flexibility needed for efficient estimation, while their small gaps keep the exploration regret under control.

%The key idea is to learn \emph{differences} of CDF values rather than CDF values individually. This shift in paradigma was first introduced by \cite{CLM26b}, and it permits the algorithm to share information across price pairs. 

%However, after elimination, the regular grid structure that supports the known efficient difference-estimation procedures is no longer available.

%Our main technical contribution is a procedure that bypasses eliminated grid points while preserving the structural features needed for efficient learning.

\subsubsection{Conservative Successive Elimination}

%The existence of suitable mappings creates a trade-off. Keeping more points queryable makes it easier to construct maps that change rarely and induce little probability overlap. On the other hand, querying a point with a large suboptimality gap produces large regret.

We implement successive elimination using two thresholds on estimated suboptimality gaps. In phase $i$, the smaller threshold defines the target set, while the larger threshold defines the queryable set. Both thresholds decrease with the phase precision $\varepsilon_i$, and their separation creates a buffer between the target points and those excluded from sampling.
In each phase we hence have:
\begin{itemize}
    \item a \emph{target set}, containing the potentially optimal points whose reward estimates must be refined; and
    \item a larger \emph{queryable set}, which also retains auxiliary price pairs whose estimated suboptimality gaps are at most a suitable multiple of the phase precision $\varepsilon_i$.
\end{itemize}

The refinement procedure may query any point in the queryable set, but it only needs to refine the estimates of the target points. These refined estimates determine which points are retained for the next phase. Auxiliary query points therefore need not have their estimates refined or be retained at the next precision.

The thresholds are chosen to preserve two guarantees: the target set contains the optimal points, and every point in the queryable set has suboptimality gap $\widetilde { \mathcal O}(\varepsilon_i)$. The first ensures that optimal actions are preserved
across phases, while the second bounds the regret
incurred by each query.

We remark that the separation between target and queryable
sets yields a generalization of the standard
successive-elimination framework that may be of independent
interest. The framework applies to any learning problem in
which estimates over a target set can be efficiently refined
using a larger set of low-regret query actions.

%Successive elimination in bilateral trade creates a tension between efficient information sharing and low exploration regret. Retaining additional points in the queryable set can preserve the geometric structure needed to estimate profits efficiently across actions. Eliminating suboptimal points tightens the bound on regret per query, but can disrupt this structure and increase the number of samples needed to refine the estimates.

%We resolve this tension through a conservative elimination rule. In phase $s$, the algorithm maintains two nested sets:
%\begin{itemize}
%    \item a \emph{target set}, containing the potentially near-optimal points whose reward estimates must be refined
%    \item a larger \emph{queryable set}, which also retains auxiliary price pairs whose estimated suboptimality gaps are at most a suitable multiple of the phase precision $\varepsilon_s$.
%\end{itemize}

%Thus, some points that are already known to be suboptimal are retained as auxiliary query points. Their gaps are small enough that querying them costs only $O(\varepsilon_s)$ regret per sample, while their presence supplies the geometric flexibility needed to efficiently refine the estimates of the points in the target set.

%This is the main difference between our approach and standard successive elimination. Rather than eliminating every certifiably suboptimal point immediately, we retain a controlled buffer of near-optimal points. The buffer makes it possible to learn the relative reward structure of the target set efficiently, and its cost is controlled at the scale of the current phase.

\subsubsection{Missing Grid Points and Relative Learning}

With the generalized successive-elimination framework in place,
the remaining challenge is to design a refinement procedure at precision $\epsilon_i$ with sample complexity  $\widetilde{\mathcal O}(\epsilon_i^{-3})$.  
%On a one-dimensional grid, estimating the CDF at each point independently is sufficiently efficient. On a two-dimensional grid, however, the same approach requires too many samples.

The main difficulty in adapting the procedure
of~\cite{castiglioni2026sample} to our setting is that
some of the grid points required by the procedure may
lie outside the queryable set.
Although conservative
successive elimination mitigates this issue by retaining
auxiliary query points, it does not guarantee that all
the points needed by the original procedure remain
available.

To address this difficulty, we build on the
\emph{relative-learning} approach introduced
by~\cite{CLM26b}. The key idea is to estimate differences between CDF values along each row, rather than estimating the absolute CDF value at every point. The absolute values can then be recovered from a single directly estimated reference point per row. Since there are only $K$ reference points, estimating them independently is feasible within our sampling budget. The remaining challenge is to efficiently learn the differences between CDF values along each row.

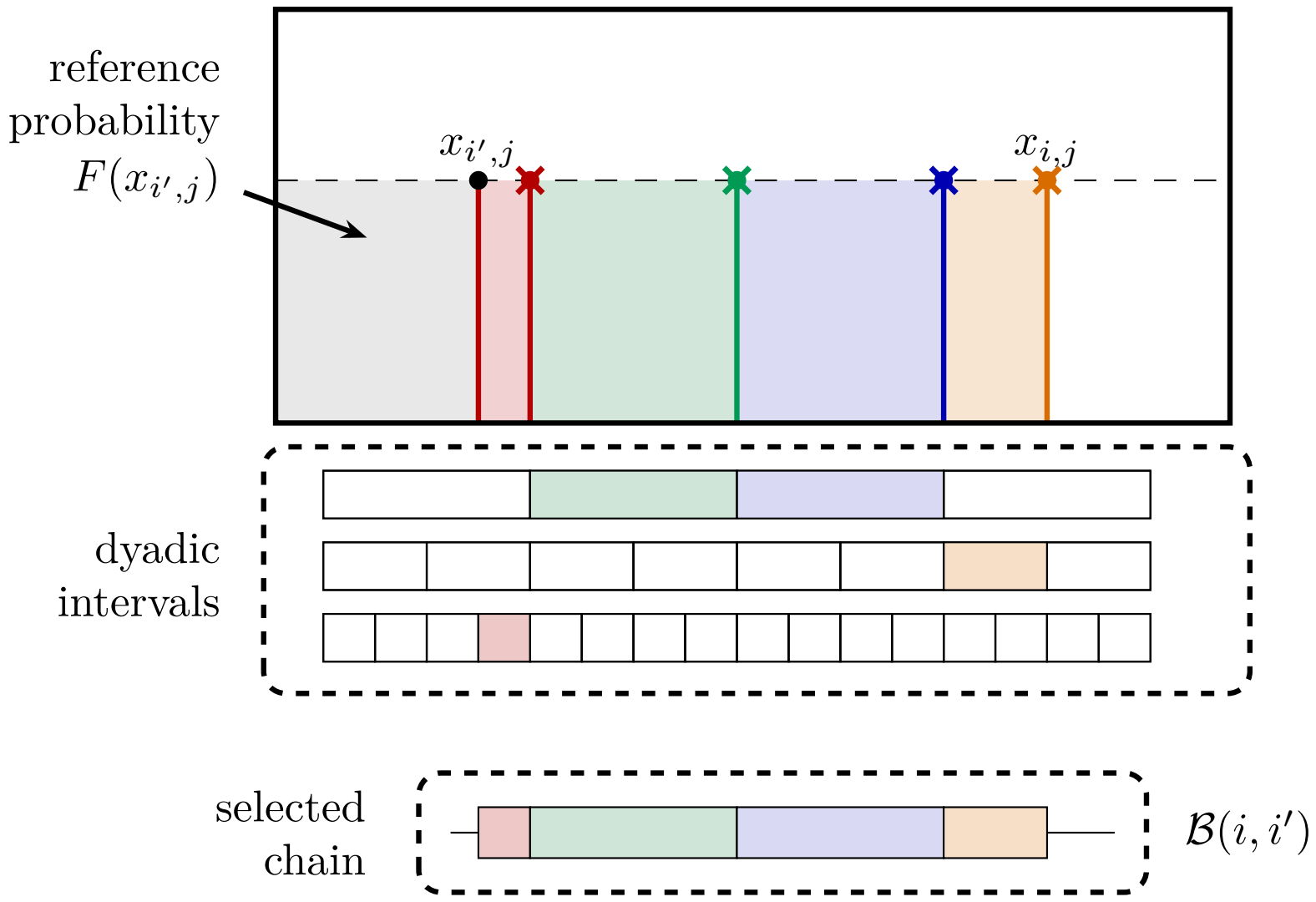
\begin{figure}[t]
\centering
\begin{tikzpicture}[x=.63cm,y=.63cm,>=Stealth,font=\small]
  \colorlet{mainblue}{blue!70!black}
  \colorlet{accentorange}{orange!85!black}
  \colorlet{sitecolor}{red!70!black}
  \colorlet{supportgreen}{ForestGreen}
  \tikzset{
    frame/.style={black,very thick},
    guide/.style={black,dash pattern=on 5pt off 4pt},
    trace/.style={dash pattern=on 1.3pt off 5.2pt,line cap=round,very thick},
    cross/.style={line width=1.3pt}
  }

  % =================================================================
  % Top panel: reference-probability rectangle.
  % =================================================================

  % The probability regions lie below row j; the upper half remains unfilled.
  \fill[gray!18] (1.2,6.25) rectangle (3.75,9.3);
  \fill[sitecolor!18] (3.75,6.25) rectangle (4.40,9.3);
  \fill[supportgreen!17] (4.40,6.25) rectangle (7.00,9.3);
  \fill[mainblue!14] (7.00,6.25) rectangle (9.60,9.3);
  \fill[accentorange!18] (9.60,6.25) rectangle (10.90,9.3);
  \draw[sitecolor,very thick] (3.75,6.25)--(3.75,9.3) (4.40,6.25)--(4.40,9.3);
  \draw[supportgreen,very thick] (7.00,6.25)--(7.00,9.3);
  \draw[mainblue,very thick] (9.60,6.25)--(9.60,9.3);
  \draw[accentorange,very thick] (10.90,6.25)--(10.90,9.3);

 %\node[font=\normalsize\bfseries] at (7.2,12.25)
  %  {Reference probability and its dyadic decomposition};
  \draw[frame] (1.2,6.25) rectangle (13.2,11.45);
  \draw[guide] (1.2,9.3)--(13.2,9.3);

  \node[anchor=east,align=right] at (.72,10.35) {reference\\probability};
  \draw[black,very thick,-{Stealth[length=2mm]}] (.8,9.15)--(2.35,8.58);
  \node[anchor=east] at (.72,9.3) { $F(x_{i',j})$};

  % Query/reference sites at row j.
  \foreach \x/\col in {3.75/black,4.40/sitecolor,7.00/supportgreen,9.60/mainblue,10.90/accentorange} {
    \fill[\col] (\x,9.3) circle (2.1pt);
  }
  \draw[sitecolor,cross] (4.40-.16,9.3-.16)--(4.40+.16,9.3+.16)
                         (4.40-.16,9.3+.16)--(4.40+.16,9.3-.16);
  \draw[supportgreen,cross] (7.00-.16,9.3-.16)--(7.00+.16,9.3+.16)
                             (7.00-.16,9.3+.16)--(7.00+.16,9.3-.16);
  \draw[mainblue,cross] (9.60-.16,9.3-.16)--(9.60+.16,9.3+.16)
                         (9.60-.16,9.3+.16)--(9.60+.16,9.3-.16);
  \draw[accentorange,cross] (10.90-.16,9.3-.16)--(10.90+.16,9.3+.16)
                              (10.90-.16,9.3+.16)--(10.90+.16,9.3-.16);
  \node[black,below=-14pt] at (3.75,9.3) {$x_{i',j}$};
  \node[black,below=-14pt] at (10.90,9.3) {$x_{i,j}$};

  % Traces from the reference region to the data structure.
  %\draw[sitecolor,trace] (3.75,9.05)--(3.75,3.25);
  %\draw[sitecolor,trace] (4.40,9.05)--(4.40,3.15);
  %\draw[supportgreen,trace] (7.00,9.05)--(7.00,3.8);
  %\draw[mainblue,trace] (9.60,9.05)--(9.60,3.8);
  %\draw[accentorange,trace] (10.90,9.05)--(10.90,3.8);

  % =================================================================
  % Middle panel: dyadic interval hierarchy.
  % =================================================================
  \draw[black,dashed,very thick,rounded corners=5pt] (1.05,2.85) rectangle (13.45,5.95);
  \node[anchor=east,align=right] at (.72,4.35) {dyadic\\intervals};
  % Three distinct scales make the dyadic-tree structure explicit.
  \foreach \a/\b in {1.80/2.45,2.45/3.10,3.10/3.75,3.75/4.40,4.40/5.05,5.05/5.70,5.70/6.35,6.35/7.00,7.00/7.65,7.65/8.30,8.30/8.95,8.95/9.60,9.60/10.25,10.25/10.90,10.90/11.55,11.55/12.20}
    {\draw[black] (\a,3.25) rectangle (\b,3.85);}
  \foreach \a/\b in {1.80/3.10,3.10/4.40,4.40/5.70,5.70/7.00,7.00/8.30,8.30/9.60,9.60/10.90,10.90/12.20}
    {\draw[black] (\a,4.15) rectangle (\b,4.75);}
  \foreach \a/\b in {1.80/4.40,4.40/7.00,7.00/9.60,9.60/12.20}
    {\draw[black] (\a,5.05) rectangle (\b,5.65);}
  % Colored fills mark the selected intervals; neutral outlines retain the tree.
  \fill[sitecolor!22] (3.75,3.25) rectangle (4.40,3.85);
  \fill[supportgreen!18] (4.40,5.05) rectangle (7.00,5.65);
  \fill[mainblue!15] (7.00,5.05) rectangle (9.60,5.65);
  \fill[accentorange!22] (9.60,4.15) rectangle (10.90,4.75);
  \foreach \a/\b in {1.80/2.45,2.45/3.10,3.10/3.75,3.75/4.40,4.40/5.05,5.05/5.70,5.70/6.35,6.35/7.00,7.00/7.65,7.65/8.30,8.30/8.95,8.95/9.60,9.60/10.25,10.25/10.90,10.90/11.55,11.55/12.20}
    {\draw[black] (\a,3.25) rectangle (\b,3.85);}
  \foreach \a/\b in {1.80/3.10,3.10/4.40,4.40/5.70,5.70/7.00,7.00/8.30,8.30/9.60,9.60/10.90,10.90/12.20}
    {\draw[black] (\a,4.15) rectangle (\b,4.75);}
  \foreach \a/\b in {1.80/4.40,4.40/7.00,7.00/9.60,9.60/12.20}
    {\draw[black] (\a,5.05) rectangle (\b,5.65);}

  % =================================================================
  % Bottom panel: the selected chain.
  % =================================================================
  \draw[black,dashed,very thick,rounded corners=5pt] (3.0,.35) rectangle (12.15,1.85);
  \draw[black] (3.4,1.1)--(11.75,1.1);
  \filldraw[fill=sitecolor!22] (3.75,.78) rectangle (4.40,1.42);
  \filldraw[fill=supportgreen!18] (4.40,.78) rectangle (7.00,1.42);
  \filldraw[fill=mainblue!15] (7.00,.78) rectangle (9.60,1.42);
  \filldraw[fill=accentorange!22] (9.60,.78) rectangle (10.90,1.42);
  \node[anchor=east,align=right] at (2.55,1.1) {selected\\chain};
  \node[anchor=west] at (12.42,1.1) {$\mathcal B(i,i')$};
\end{tikzpicture}
\caption{A decomposition of the region between $x_{i,j}$ and $x_{i',j}$. Top: the probability region associated with the grid point $x_{i',j}$ is gray, while the union of the colored regions represents the probability regions associated with the difference between $x_{i,j}$ and $x_{i',j}$. Middle: the dyadic interval tree of cardinality $\mathcal{O}(K)$ used to decompose the differences. Bottom: the selected intervals form the consecutive chain $\mathcal B(i,i')$ used to reconstruct the CDF difference. Colors identify the same interval throughout.}
\label{fig:dyadic-reference-decomposition}
\end{figure}

To illustrate how these differences can be learned efficiently,
we first consider the simpler setting in which the entire grid
is queryable. The key tool is a dyadic decomposition.
Let $x_{i,j}$ denote the grid point in column $i$ and row $j$. For two points $x_{i,j}$ and $x_{i',j}$ on the same row, their CDF difference can be written as
 
\begin{align}
D_j(i,i')
&=F(x_{i,j})-F(x_{i',j}) =\sum_{(a,b)\in\mathcal{B}(i,i')}
\bigl(F(x_{b,j})-F(x_{a,j})\bigr).
\label{eq: naive decomp}
\end{align}
Here, $\mathcal{B}(i,i')$ denotes a canonical dyadic
decomposition of the interval $(i',i]$ into
$\mathcal{O}(\log K)$ dyadic intervals. All intervals appearing in these decompositions belong
to a common family $\mathcal I^\star$ of size
$\mathcal{O}(K)$. We refer the reader to Section~\ref{sec:dyadic}
for the formal definitions and to
Figure~\ref{fig:dyadic-reference-decomposition}
for an illustration.

This representation allows comparisons between different pairs of points to share the same elementary differences. For each row, learning all $\mathcal O(K^2)$ pairwise differences therefore reduces to learning the $\mathcal O(K)$ differences associated with $\mathcal I^\star$. 
Moreover, on a complete grid, a one-dimensional binary-search
procedure can estimate the differences associated with a fixed
dyadic interval across all rows to additive precision
$\varepsilon_i$, using
$\widetilde{\mathcal O}(\varepsilon_i^{-3})$
samples~\cite{castiglioni2026sample}.

%This representation compresses the learning problem. Rather than estimating all $\mathcal{O}(K^2)$ possible horizontal differences separately, it is enough to estimate the $\mathcal{O}(K)$ elementary dyadic differences in $\mathcal{I}^{\star}$. On a complete grid, the differences associated with one dyadic interval can be estimated simultaneously over all rows using a one-dimensional binary-search procedure (\cite{castiglioni2026sample}).

Successive elimination, however, introduces holes into the grid, breaking the required regularity. Even when both points being compared belong to the target set, an intermediate endpoint in their dyadic decomposition may lie outside the queryable set. Applying the decomposition in Eq.~\eqref{eq: naive decomp} unchanged can therefore require queries at price pairs whose suboptimality gaps are too large for the current phase.

\subsubsection{Rerouting Dyadic Decompositions}

To overcome this issue, we  reroute the dyadic decomposition through queryable representatives. For each row $j$, we construct a mapping
\[
v_j:\{0,\ldots,K-1\}\to\{0,\ldots,K-1\},
\]
where $v_j(i)$ denotes the mapped representative index associated
with $i$, and $x_{v_j(i),j}$ its mapped representative. By composing this rerouting with the dyadic decomposition, we get:   
\begin{align}
D_j(i,i')
&=F(x_{i,j})-F(x_{i',j}) \nonumber\\
&\hspace{-0.5cm}=\bigl(F(x_{i,j})-F(x_{v_j(i),j})\bigr)
+\sum_{(a,b)\in\mathcal{B}(i,i')}
\bigl(F(x_{v_j(b),j})-F(x_{v_j(a),j})\bigr) +\bigl(F(x_{v_j(i'),j})-F(x_{i',j})\bigr).
\label{eq: map decomp}
\end{align}

The middle sum allows the same elementary differences to be reused across different comparisons. Figure~\ref{fig:dyadic-mapped-jumps} illustrates an example of the dyadic decomposition using mapped representatives. The main challenge is to construct the mappings so that all
differences required to compare target points can be
estimated efficiently using only queryable grid points.

To this end, we identify three properties that the mappings must satisfy.

\paragraph{Queryability.}

Consider two target points $x_{i,j}$ and $x_{i',j}$
on the same row. Their CDF difference is expressed
in~\eqref{eq: map decomp} as a sum of elementary
differences and two correction terms.
To estimate these differences, our procedure queries
the grid points corresponding to their endpoints.
We therefore require that every difference
appearing in the decomposition of two target points
has both endpoints in the queryable set. Since the points in this set have small gaps, this condition bounds the regret incurred by each query used to estimate the mapped elementary differences.

\begin{requirement}[Queryability]\label{req: 1}
For every row $j\in\{0,\ldots,K-1\}$ and every pair of target points $x_{i,j},x_{i',j}$, all terms in Equation~\eqref{eq: map decomp} must involve only
points belonging to the queryable set of the
current phase.
\end{requirement}

\begin{figure}[t]
\centering
\begin{tikzpicture}[x=0.49cm,y=0.49cm,>=Stealth,font=\small]

\begin{scope}
  \node[font=\normalsize\bfseries] at (4,9.5)
    {Mapped representative index paths};
  \draw[step=1,gray!25] (0,0) grid (8,8);
  \draw[->] (0,0)--(8.6,0) node[right]{column};
  \draw[->] (0,0)--(0,8.6) node[left]{row};
  \foreach \i in {0,2,4,6,8} {
    \node[below] at (\i,0){\i};
    \node[left] at (0,\i){\i};
  }

  % The two mapped endpoints change on different row transitions.
  \draw[very thick,blue!65!black]
    (2,0)--(2,1.5)--(3,1.5)--(3,4.5)--(1,4.5)--(1,8);
  \draw[very thick,orange!80!black]
    (4,0)--(4,2.5)--(6,2.5)--(6,5.5)--(5,5.5)--(5,8);

  % Discrete row positions of v_j(2).
  \foreach \j in {0,1}
    \fill[blue!65!black] (2,\j) circle(2pt);
  \foreach \j in {2,3,4}
    \fill[blue!65!black] (3,\j) circle(2pt);
  \foreach \j in {5,6,7,8}
    \fill[blue!65!black] (1,\j) circle(2pt);

  % Discrete row positions of v_j(4).
  \foreach \j in {0,1,2}
    \fill[orange!80!black] (4,\j) circle(2pt);
  \foreach \j in {3,4,5}
    \fill[orange!80!black] (6,\j) circle(2pt);
  \foreach \j in {6,7,8}
    \fill[orange!80!black] (5,\j) circle(2pt);

  % Previous representatives at the four distinct changes.
  \foreach \x/\y in {2/2,4/3,3/5,6/6}
    \draw[red!70!black,thick]
      (\x-0.13,\y-0.13)--(\x+0.13,\y+0.13)
      (\x-0.13,\y+0.13)--(\x+0.13,\y-0.13);

  \node[blue!65!black,right] at (1,7.5){$v_j(2)$};
  \node[orange!80!black,right] at (5,7.5){$v_j(4)$};
  \node[align=center,font=\normalsize] at (4,-1.8)
    {One shared path per dyadic junction};
\end{scope}

\begin{scope}[xshift=7.7cm]
  \node[font=\normalsize\bfseries] at (4,9.5)
    {Fixed-endpoint blocks};
  \draw[step=1,gray!25] (0,0) grid (8,8);
  \draw[->] (0,0)--(8.6,0) node[right]{column};
  \draw[->] (0,0)--(0,8.6) node[left]{row};
  \foreach \i in {0,2,4,6,8} {
    \node[below] at (\i,0){\i};
    \node[left] at (0,\i){\i};
  }

  % A new block starts whenever either endpoint changes.
  \filldraw[fill=blue!12,draw=blue!65!black] (2,0) rectangle (4,1);
  \filldraw[fill=orange!12,draw=orange!80!black] (3,2) rectangle (4,2);
  \filldraw[fill=blue!12,draw=blue!65!black] (3,3) rectangle (6,4);
  \filldraw[fill=orange!12,draw=orange!80!black] (1,5) rectangle (6,5);
  \filldraw[fill=blue!12,draw=blue!65!black] (1,6) rectangle (5,8);

  \node at (3,.5){1};
  \node at (3.5,2){2};
  \node at (4.5,3.5){3};
  \node at (3.5,5){4};
  \node at (3,7){5};

  \node[align=center,font=\normalsize] at (4,-1.8)
    {Disjoint row spans within a mapped column};
\end{scope}

\end{tikzpicture}

\caption{The dyadic mapped column associated with $(2,4]$ over rows
$0,\ldots,8$. The left mapped representative index is \(2,3,1\) on row ranges
\(0\)--\(1\), \(2\)--\(4\), and \(5\)--\(8\), respectively; the right mapped
representative index is \(4,6,5\) on row ranges \(0\)--\(2\), \(3\)--\(5\), and
\(6\)--\(8\). Left: the two mapped representative index paths change at different row
transitions while remaining in their permitted intervals
\(I_2=[0,4]\) and \(I_4=[0,8]\). Crosses mark previous mapped representatives
that are no longer queryable. Right: each change of either endpoint starts
a new fixed-endpoint block; the rectangles are half-open and may overlap
horizontally.}
\label{fig:bent}
\end{figure}
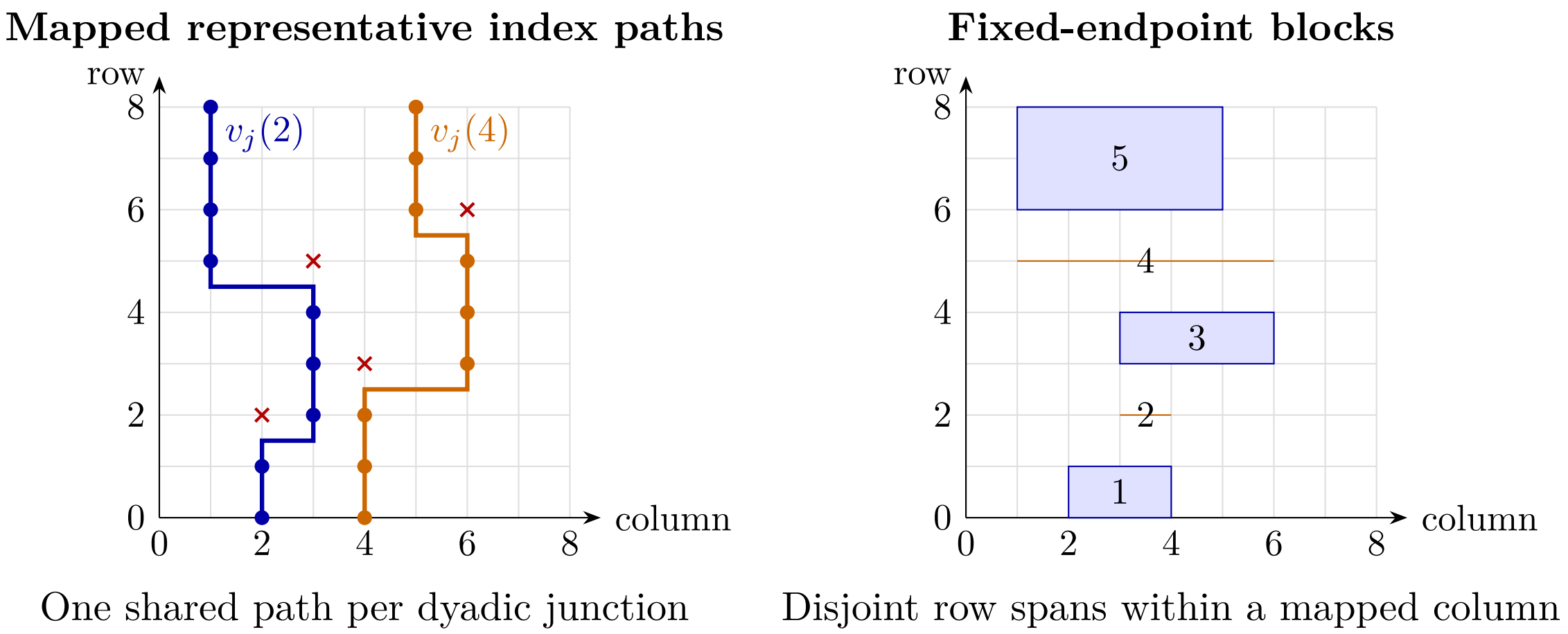

\paragraph{Few binary searches.} 
For a fixed dyadic pair $(i,i')\in\mathcal I^\star$, the mapped representative indices $v_j(i)$ and $v_j(i')$ may vary across rows.
Consequently, the region between
the corresponding grid points need not form a rectangle,
as illustrated in Figure~\ref{fig:bent}.
To address this, we partition the rows into maximal
fixed-endpoint blocks on which both mapped representative indices remain
fixed. Within each such block, we can apply the binary-search procedure.

However, each block requires a separate initialization,
which contributes to the sample complexity. For each pair in the dyadic decomposition, there is one initial block, and a new block begins when at least one
of its mapped representatives changes between consecutive rows. Thus, if $C$ denotes the total number of changes  across all rows, the total number of blocks is of order $|\mathcal I^\star|+C$. Since $|\mathcal I^\star|=\mathcal O(K)$, the initial
blocks are affordable within our sampling budget.
The challenge is therefore to control the additional
initialization cost caused by changes in the mappings.

\begin{requirement}[Few binary searches]\label{req: 2}
At precision $\varepsilon_i$, the mappings $\{v_j(\cdot)\}_{j\in \{0,\ldots,K-1\}}$ satisfy
\[
    C :=\sum_{j=0}^{K-2}\Bigg(\sum_{(a,b)\in\Dyad}\ind\left\{(v_j(a),v_j(b))\neq(v_{j+1}(a),v_{j+1}(b))\right\}+\sum_{i=0}^{K-1}\ind\left\{v_j(i)\neq v_{j+1}(i)\right\}\Bigg)
    \leq \widetilde{\mathcal{O}}\left(\frac{1}{\varepsilon_i}\right).
\]
\end{requirement}

\paragraph{Low probability overlap.}

Controlling the number of blocks bounds the number of binary searches, but the cost of each binary search also depends on the probability mass of its rectangular region. The search recursively subdivides regions whose mass exceeds the current precision scale $\varepsilon_i$. Beyond initialization, the number of rows tested is therefore bounded by the probability mass of the rectangle divided by $\varepsilon_i$.

Ideally, the rectangles would be disjoint, so that each unit of probability mass would contribute to the search cost only once. The mapped rectangles may overlap, however, causing the same mass to be charged repeatedly across different searches. We therefore control their cumulative probability mass, counted with multiplicity.

\begin{requirement}[Low probability overlap]\label{req: 3}
Let $\mathcal V$ denote the cumulative probability mass of all blocks to estimate. The mappings $\{v_j(\cdot)\}_{j\in\{0,\ldots,K-1\}}$ must ensure that this sum is at most polylogarithmic in $K$ and $1/\varepsilon_i$, namely,
\[
\mathcal V\leq \textnormal{poly} \log(K/\varepsilon_i). 
\]

\end{requirement}

\subsubsection{Putting Everything Together}

We can now summarize how these properties
yield the desired regret.

The refinement procedure incurs two types of sampling cost: initializing each fixed-endpoint block and testing additional rows during binary search. The number of blocks controls the first cost, while their cumulative probability mass controls the second. Each directly estimated CDF value requires $\widetilde{\mathcal{O}}(\varepsilon_i^{-2})$ samples to achieve precision $\varepsilon_i$. Combining the bound \(\mathcal O(K+C)\) on the number of blocks with Requirements~\ref{req: 2} and~\ref{req: 3}, we obtain the following overall sample-complexity bound: 
\[
\widetilde{\mathcal O}\left(
\frac{1}{\varepsilon^2_i} \left(K+ C
+\frac{\mathcal V}{\varepsilon_i}\right)
\right)
=
\widetilde{\mathcal O}\left(
\frac{K}{\varepsilon^2_i}
+\frac{1}{\varepsilon^3_i}
\right),
\]

Estimating one point per row then converts the relative estimates into absolute CDF estimates, without changing the order of this bound.

The missing piece is the regret incurred while gathering these samples. Thanks to Requirement~\ref{req: 1}, every required query lies in the queryable set, whose points have suboptimality gaps $\widetilde{\mathcal{O}}(\varepsilon_i)$. Multiplying the sample bound by this per-query gap gives a per-phase regret of $\widetilde{\mathcal{O}}(K/\varepsilon_i+\varepsilon_i^{-2})$. %Since the phase precisions decrease geometrically, the total exploration regret is dominated by the final phase.
Choosing $K=\Theta(T^{1/3})$ and a final precision at scale $T^{-1/3}$ then leads to the desired $\widetilde{\mathcal{O}}(T^{2/3})$ regret rate.

The preceding argument shows that the desired regret bound
follows from the three properties of the mappings.
However, it remains to establish that these properties
can be satisfied simultaneously. This is the central
technical challenge of our approach: the mappings must
ensure queryability of the required differences while
limiting both the number of changes across rows and the
cumulative probability mass. We next explain how we construct mappings satisfying all three requirements.

\subsubsection{Constructing the Mappings}

The construction follows three main principles that allow us
to satisfy the three requirements simultaneously.

\begin{figure}
    \centering
    \begin{tikzpicture}[x=.60cm,y=1cm,>=Stealth,font=\small,
  target/.style={diamond,fill=accentorange,draw=white,
    line width=.4pt,inner sep=1.8pt},
  junction/.style={circle,fill=black,inner sep=1.6pt},
  window/.style={draw=mainblue!65,line width=.6pt,fill=mainblue!7},
  guide/.style={gray!22,line width=.3pt},
  panel/.style={anchor=west,font=\small\bfseries},
  note/.style={anchor=west,align=left,font=\footnotesize},
  region/.style={line width=.65pt,fill opacity=.13}]
\colorlet{mainblue}{blue!70!black}
\colorlet{accentorange}{orange!85!black}
\colorlet{sitecolor}{red!70!black}
\colorlet{supportgreen}{green!50!black}

% (a) All three window bars refer to the SAME row j.
\node[panel] at (-.6,1.12)
  {(a) Permitted intervals contain a target endpoint};
%\node[anchor=east,font=\footnotesize] at (25,1.12) {$K=17$};
\foreach \i in {0,...,16}
  \draw[guide] (\i,-2.92)--(\i,.12);
\draw[gray!55] (0,0)--(16,0);
\foreach \i in {0,...,16} {
  \fill[gray!60] (\i,0) circle (.8pt);
  \node[below=4pt,font=\scriptsize] at (\i,0) {$\i$};
}
\node[anchor=east,align=right] at (-.35,0) {row $j$};

% Canonical intervals (3,4], (4,8], (8,12], (12,14].
\foreach \a/\b/\col in {3/4/sitecolor,4/8/supportgreen,
                         8/12/mainblue,12/14/accentorange} {
  \draw[gray!65,line width=1.1pt] (\a,0)--(\b,0);
  \draw[gray!65,line width=.8pt]
    (\a,.06) .. controls (\a,.47) and (\b,.47) .. (\b,.06);
}
\foreach \i/\col in {4/sitecolor,8/supportgreen,12/mainblue}
  \fill[\col] (\i,0) circle (2pt);
\node[target] at (3,0) {};
\node[target] at (14,0) {};
\node[above,accentorange] at (3,.45) {$x_{3,j}$};
\node[above,accentorange] at (14,.45) {$x_{14,j}$};

% Discrete windows are drawn as their enclosing bars.
\foreach \a/\b/\y/\col in {0/8/-1.05/sitecolor,0/16/-1.85/supportgreen,8/16/-2.65/mainblue}
  \draw[window,draw=\col,fill=\col!9] (\a,\y-.19) rectangle (\b,\y+.19);
\node[anchor=west] at (16.45,-1.05) {$v_j(4)\in I_4=\{0,\ldots,8\}$};
\node[anchor=west] at (16.45,-1.85) {$v_j(8)\in I_8=\{0,\ldots,16\}$};
\node[anchor=west] at (16.45,-2.65) {$v_j(12)\in I_{12}=\{8,\ldots,16\}$};

% Only endpoints belonging to each window are copied into its bar.
\draw[accentorange!55,densely dashed] (3,-.38)--(3,-1.85);
\draw[accentorange!55,densely dashed] (14,-.38)--(14,-2.65);
\foreach \x/\y in {3/-1.05,3/-1.85,14/-1.85,14/-2.65}
  \node[target] at (\x,\y) {};

% Fixed theoretical junctions: the midpoints of the selection windows.
% No particular value of the mapping v_j is depicted.
\foreach \k/\y/\col in {4/-1.05/sitecolor,8/-1.85/supportgreen,12/-2.65/mainblue} {
  \node[junction,fill=\col] at (\k,\y) {};
  \node[above=6pt,font=\footnotesize] at (\k,\y) {$\k$};
}

% (b) Fix h=4. Both panels refer to the same row j.
% Neighbors are endpoints of dyadic pairs, not all indices in I_i.
\begin{scope}[yshift=-4.85cm]
\node[panel] at (-.6,.95)
  {(b) Nested windows at a fixed scale};
%\node[anchor=east,font=\footnotesize] at (25,.95) {$h=4$};
\draw[gray!55] (0,0)--(16,0);
\foreach \i in {0,...,16} {
  \fill[gray!60] (\i,0) circle (.8pt);
  \node[below=4pt,font=\scriptsize] at (\i,0) {$\i$};
}
\node[anchor=east] at (-.35,0) {row $j$};
% Exact requested tints: 100% at h_i=4, 60% at h_i=2, 30% at h_i=1.
\foreach \i/\col in {4/sitecolor,12/mainblue} {
  \fill[\col] (\i,0) circle (2.5pt);
  \node[above=6pt,\col,font=\footnotesize] at (\i,0) {$h_{\i}=4$};
}
\foreach \i/\col in {2/sitecolor,6/sitecolor,10/mainblue,14/mainblue}
  \fill[\col!60] (\i,0) circle (2.5pt);
\foreach \i/\col in {3/sitecolor,5/sitecolor,11/mainblue,13/mainblue}
  \fill[\col!30] (\i,0) circle (2.5pt);

% Color legend: the two families share the same intensity scale.
\foreach \y/\tint/\h in {0/100/4,-.45/60/2,-.9/30/1} {
  \fill[sitecolor!\tint] (17,\y) circle (2.5pt);
  \fill[mainblue!\tint] (17.65,\y) circle (2.5pt);
  \node[anchor=west,font=\footnotesize] at (18.15,\y) {$h_i=\h$};
}

\begin{scope}[yshift=-.3cm]
% Parent windows: colored regions, without a background coordinate plane.
\foreach \a/\b/\i/\col in {0/8/4/sitecolor,8/16/12/mainblue} {
  \draw[draw=\col,fill=\col!5,line width=.7pt]
    (\a,-3.64) rectangle (\b,-.82);
  \node[above,font=\footnotesize,\col] at (\i,-.82)
    {$I_{\i}=\{\a,\ldots,\b\}$};
}

% All lower-scale dyadic neighbors of 4 and 12, with their exact windows.
% Four separate levels make every containment visible.
\foreach \a/\b/\i/\y/\tint/\wash/\col in {
  0/4/2/-1.38/60/15/sitecolor,
  4/8/6/-2.03/60/15/sitecolor,
  2/4/3/-2.68/30/8/sitecolor,
  4/6/5/-3.33/30/8/sitecolor,
  8/12/10/-1.38/60/15/mainblue,
  12/16/14/-2.03/60/15/mainblue,
  10/12/11/-2.68/30/8/mainblue,
  12/14/13/-3.33/30/8/mainblue} {
  \draw[draw=\col!\tint,fill=\col!\wash,line width=.9pt]
    (\a,\y-.12) rectangle (\b,\y+.12);
  \fill[\col!\tint] (\i,\y) circle (2pt);
  \node[above=5pt,font=\footnotesize] at (\i,\y) {$I_{\i}$};
}
\node[anchor=west,align=left,font=\footnotesize] at (16.45,-2.25)
  {for all $(i,i')$ or $(i',i)$ in $\Dyad$ if \\[4pt]
   $h_{i'}<h_i,$ then $I_{i'}\subseteq I_i$\\[4pt]
    $v_j(i),v_j(i')\in I_i$};

% Physical supports corresponding to the two parent windows.
\draw[sitecolor,line width=1.4pt] (0,-4.12)--(8,-4.12);
\draw[mainblue,line width=1.4pt] (8,-4.12)--(16,-4.12);
\draw[sitecolor,fill=white,line width=.8pt] (0,-4.12) circle (1.7pt);
\fill[sitecolor] (8,-4.12) circle (1.7pt);
\fill[mainblue] (16,-4.12) circle (1.7pt);
\node[below=5pt,font=\footnotesize] at (4,-4.12) {$(z_0,z_8]$};
\node[below=5pt,font=\footnotesize] at (12,-4.12) {$(z_8,z_{16}]$};
\node[anchor=west,align=left,font=\footnotesize] at (16.45,-4.12)
  {disjoint intervals $(z_{i-h_i},z_{i+h_i}]$\\(half-open boundaries)};
\end{scope}
\end{scope}
\end{tikzpicture}
    \caption{Permitted intervals and their dyadic structure. (a) In the canonical dyadic decomposition between the target points $x_{3,j}$ and $x_{14,j}$, each intermediate junction $i\in\{4,8,12\}$ has a permitted interval $I_i$ containing at least one of the two target points. (b) Permitted intervals are nested across dyadic scales: for every $(a,b)\in\mathcal I^\star$, the interval between the mapped representative indices $v_j(a)$ and $v_j(b)$ is contained in the permitted interval of the endpoint with the larger permitted radius. At each scale, the corresponding half-open intervals $(z_{i-h_i},z_{i+h_i}]$ are disjoint, as illustrated by $(z_0,z_8]$ and $(z_8,z_{16}]$.
    }
    \label{fig: qry e prob overlap}
\end{figure}
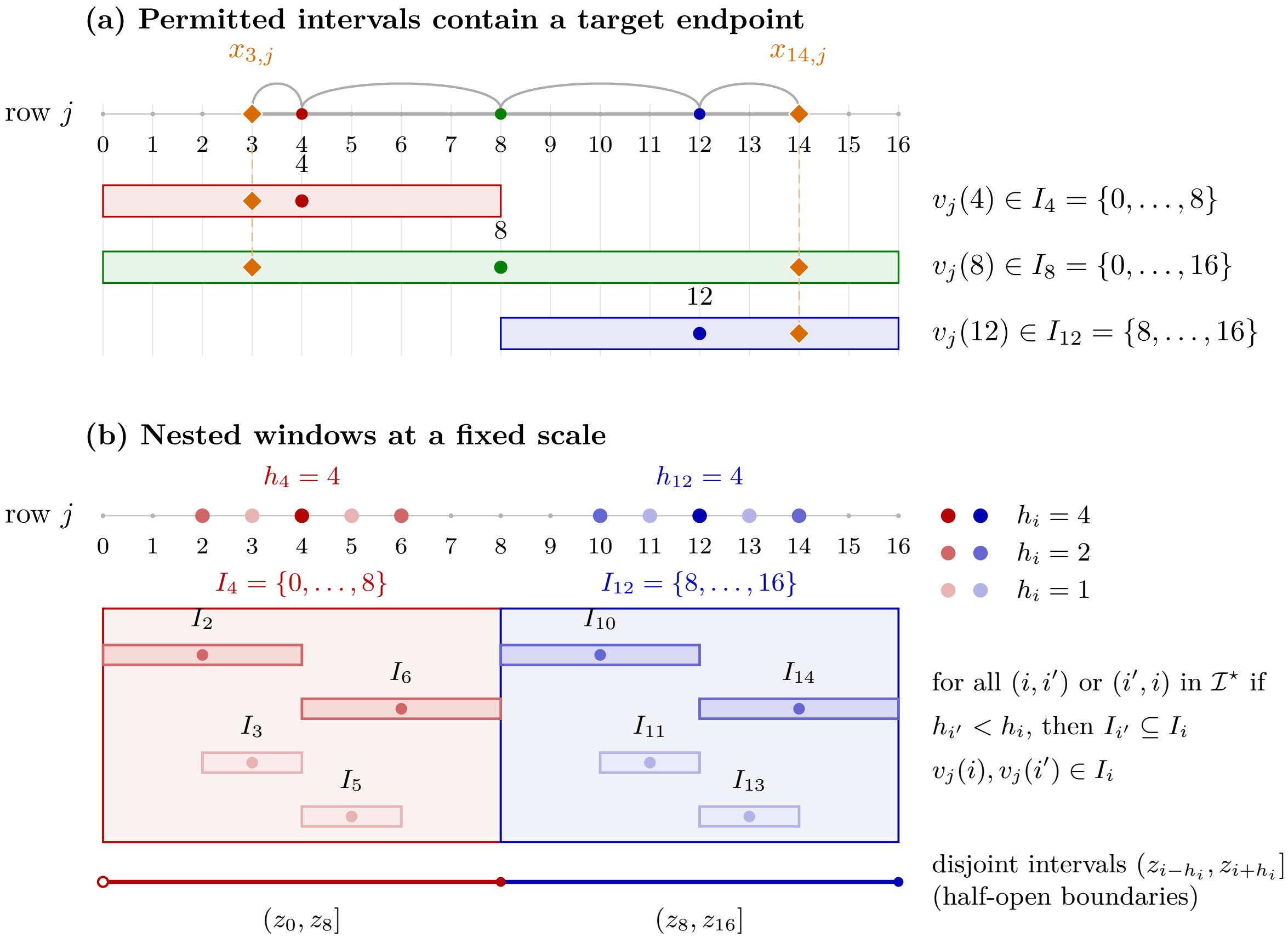

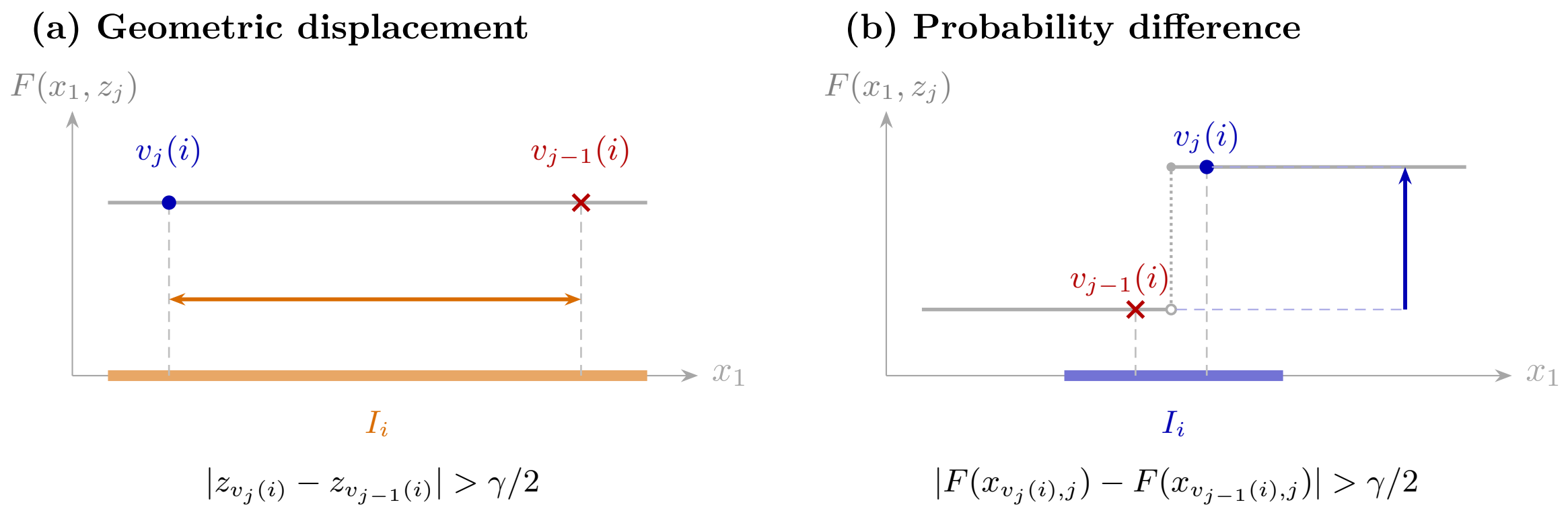
\begin{figure}
    \centering
    \begin{tikzpicture}[x=1cm,y=1cm,>=Stealth,font=\small,
  panel/.style={anchor=west,font=\small\bfseries},
  guide/.style={gray!50,densely dashed,line width=.5pt}]
\colorlet{mainblue}{blue!70!black}
\colorlet{accentorange}{orange!85!black}
\colorlet{sitecolor}{red!70!black}

% (a) Leftward movement across a plateau of the CDF increases the margin.
\node[panel] at (0,4) {(a) Geometric displacement};
\draw[->,gray!70] (.55,.6)--(6.7,.6) node[right] {$x_1$};
\draw[->,gray!70] (.55,.6)--(.55,3.2);
\node[gray,above right,font=\footnotesize] at (-0.2,3.14) {$F(x_1,z_j)$};
\draw[accentorange!60,line width=3pt] (.9,.6)--(6.2,.6);
\node[below=6pt,accentorange,font=\footnotesize] at (3.55,.6) {$I_i$};
\draw[gray!65,line width=1pt] (.9,2.3)--(6.2,2.3);
\foreach \x in {1.5,5.55} \draw[guide] (\x,.6)--(\x,2.3);
\fill[mainblue] (1.5,2.3) circle (2pt);
\draw[sitecolor,line width=1pt] (5.47,2.22)--(5.63,2.38)
  (5.47,2.38)--(5.63,2.22);
\node[above=6pt,mainblue,font=\footnotesize] at (1.5,2.3) {$v_j(i)$};
\node[above=6pt,sitecolor,font=\footnotesize] at (5.55,2.3) {$v_{j-1}(i)$ };
\draw[accentorange,line width=1.1pt,{Stealth[length=1.6mm]}-{Stealth[length=1.6mm]}]
  (1.5,1.35)--(5.55,1.35);
\node[font=\footnotesize] at (3.5,-.48) {$|z_{v_{j}(i)}-z_{v_{j-1}(i)}|>\gamma/2$};

% (b) A short spatial move can cross a large CDF jump.
\node[panel] at (8,4) {(b) Probability difference};
\draw[->,gray!70] (8.55,.6)--(14.7,.6) node[right] {$x_1$};
\draw[->,gray!70] (8.55,.6)--(8.55,3.2);
\node[gray,above right,font=\footnotesize] at (7.8,3.14) {$F(x_1,z_j)$};
\draw[mainblue!55,line width=3pt] (10.3,.6)--(12.45,.6);
\node[below=6pt,mainblue,font=\footnotesize] at (11.38,.6) {$I_i$};
\draw[gray!65,line width=1pt] (8.9,1.25)--(11.35,1.25);
\draw[gray!65,densely dotted,line width=.8pt] (11.35,1.25)--(11.35,2.65);
\draw[gray!65,line width=1pt] (11.35,2.65)--(14.25,2.65);
\draw[gray!65,fill=white,line width=.7pt] (11.35,1.25) circle (1.3pt);
\fill[gray!65] (11.35,2.65) circle (1.3pt);
\draw[guide] (11,.6)--(11,1.25) (11.7,.6)--(11.7,2.65);
\draw[sitecolor,line width=1pt] (10.92,1.17)--(11.08,1.33)
  (10.92,1.33)--(11.08,1.17);
\fill[mainblue] (11.7,2.65) circle (2pt);
\node[above ,sitecolor,font=\footnotesize] at (11,1.25) {\hspace{-0.3cm}$v_{j-1}(i)$ };
\node[above ,mainblue,font=\footnotesize] at (11.7,2.65) {$v_{j}(i)$ };
\draw[mainblue!35,densely dashed] (11.75,2.65)--(13.65,2.65)
  (11.4,1.25)--(13.65,1.25);
\draw[mainblue,line width=1.1pt,-{Stealth[length=1.7mm]}] (13.65,1.25)--(13.65,2.65);
\node[font=\footnotesize] at (11.4,-.48)
  {$|F(x_{v_{j}(i),j})-F(x_{v_{j-1}(i),j})|>\gamma/2$};
\end{tikzpicture}
    \caption{A large profit improvement requires displacement or probability mass.
Changing the map for $i$ from row $j-1$ to row $j$ requires an increase in profit of $\gamma=76\varepsilon_i\ell_K$, \emph{i.e.}, replacing an unqueryable point by a target point yields
$\gamma<r(x_{v_{j}(i),j})-r(x_{v_{j-1}(i),j})\le |z_{v_{j}(i)}-z_{v_{j-1}(i)}|+|F(x_{v_{j}(i),j})-F(x_{v_{j-1}(i),j})|$.
Thus at least one of the two differences exceeds $\gamma/2$.
(a) A leftward move, in the best case scenario across a CDF plateau, improves the profit margin by the displacement length;
a large displacement requires a wide permitted interval, so fewer dyadic junctions occur at that scale.
(b) A rightward move can be shorter but must cross substantial probability mass.
The total number of displacements can be bounded through a potential argument, using the boundedness of both probability and space.}
    \label{fig: potential fun}
\end{figure}

\begin{itemize}
    \item \textbf{Consistency across rows.}
    The procedure constructs the mappings row by row, starting at $j=0$ and proceeding in increasing order. When constructing $v_{j+1}(\cdot)$, it preserves the values of $v_j(\cdot)$ wherever possible. This limits the number of changes between consecutive rows and, consequently, the number of blocks and binary-search initializations. 
    
    \item \textbf{Locality of the mapped representatives.}
    Whenever $x_{v_j(i),j+1}$ is no longer queryable, we look for a suitable replacement for $v_{j+1}(i)$. This creates a tension between Requirements~\ref{req: 1} and~\ref{req: 3}.
    On the one hand, to limit the cumulative  probability mass, we would like the new representative index $v_{j+1}(i)$ to remain as close as possible to $i$. On the other hand, guaranteeing queryability may require selecting a point farther away.

    The key observation is that the search radius needed to find a replacement depends on the role of index $i$ in the dyadic decomposition. 
    Indeed, the decomposition of a difference relies on the longest internal dyadic intervals contained in the original interval.
    Consequently, indices divisible only by small powers of two can serve as endpoints only of short dyadic intervals, whereas indices divisible by larger powers of two can also serve as endpoints of longer intervals.
    
    We exploit this structure by restricting $v_{j+1}(i)$ to a permitted interval centered at $i$, of length twice the largest power of two dividing $i$. This restriction allows us to simultaneously guarantee the queryability of the required differences and control the cumulative probability mass, thereby satisfying Requirements~\ref{req: 1} and~\ref{req: 3} (see Figure~\ref{fig: qry e prob overlap}).

    \item \textbf{Changes only when necessary.}
    When constructing $v_{j+1}(i)$, we retain the previous
representative index $v_j(i)$ whenever possible.
    If $x_{v_j(i),j+1}$ does not belong to the queryable
set, we update the mapping only if the corresponding permitted interval contains a point in the target set. In that case, we choose the index of such a point as the new mapped representative index $v_{j+1}(i)$. 
    Whenever a difference between two target points requires an intermediate point, its permitted interval contains at least one of the two target points. Consequently, if a permitted interval contains no target point, the corresponding mapping can remain unchanged, even if its mapped representative is no longer queryable, without violating Requirement~\ref{req: 1} (see Figure~\ref{fig: qry e prob overlap}).

Finally, we bound the number of representative replacements through a
potential-function argument. Each necessary update replaces
a mapped representative that is not queryable on the current row
with a point in the target set. The separation between the
profit thresholds defining the target and queryable sets
ensures that every such replacement yields a sufficiently
large increase in expected profit.

The key challenge is to show that these improvements cannot
occur too frequently. A large difference in expected profit
between two points on the same row must arise from either
a large horizontal displacement or a substantial difference
in their CDF values. Large displacements are possible only
for indices with sufficiently wide permitted intervals,
which become increasingly sparse at larger dyadic scales.
On the other hand, large CDF differences correspond to
probability mass between the two points, and the disjointness
of permitted intervals at each scale limits how often this
mass can contribute.

We formalize this intuition through a profit potential function
based on the expected profits of the mapped representatives,
whose cumulative variation across rows can be bounded.
Since every replacement induces a sufficiently large
increase in the potential, this bound limits the total
number of mapping changes, establishing
Requirement~\ref{req: 2}
(see Figure~\ref{fig: potential fun}).

   % Finally, we control the number of changes using the bounded variation of the profit function. In particular, each necessary update replaces an unqueryable reference point with a point in the target set. The separation between the profit thresholds defining the target and queryable sets ensures that every mapping replacement yields a sufficiently large improvement in expected profit. To bound the number of such replacements, we exploit the fact that a large profit difference between two points on the same row must arise either from a large horizontal displacement or from a substantial difference in their CDF values. Large displacements are possible only at points with sufficiently wide selection windows, which become increasingly sparse at larger scales. On the other hand, large CDF differences correspond to probability mass between the two points, and the disjointness of the selection intervals at each scale limits how often this mass can contribute. We capture these properties through a potential function based on expected profit, whose cumulative variation across rows is bounded. Since every replacement produces a large increase in the potential, the total number of mapping changes must be small, establishing Requirement~\ref{req: 2} (see Figure~\ref{fig: potential fun}). 

\end{itemize}

\subsection{Related Work}\label{sec:related}

Below, we summarize the lines of research most closely related to our work.

\paragraph{Profit maximization in bilateral trade}

Until recently, profit maximization in repeated bilateral trade had received comparatively little attention. \cite{di2025nearly, di2026profit} study learning general incentive-compatible and individually rational mechanisms under full feedback, establishing $\widetilde{\mathcal{O}}(\sqrt{T})$ regret in the stochastic setting and against a smoothed adversary, respectively. \cite{aggarwal2024selling} study the related problem of selling joint advertisements to two non-excludable buyers, learning general revenue-maximizing mechanisms.
Their setting presents a different learning challenge: while our main obstacle is the limited information provided by one-bit feedback, theirs is the complexity of the mechanism class.

Closer to our setting, \cite{CLM26b} obtain $\widetilde{\mathcal{O}}(T^{7/10})$ regret for a broader class of objectives involving bidimensional CDFs under one-bit feedback. Their result applies to profit maximization in repeated bilateral trade with fixed prices, improving on the previous $\widetilde{\mathcal{O}}(T^{3/4})$ bound while leaving a gap to the classical $\Omega(T^{2/3})$ lower bound for dynamic pricing \cite{kleinberg2003value}.

\paragraph{Learning to maximize GFT in bilateral trade}

Most of the literature on learning in bilateral trade focuses on maximizing gains from trade (GFT). \cite{cesa2021regret,cesa2024bilateral} initiated the study of regret minimization in repeated bilateral trade. A central difficulty is that trade/no-trade feedback does not reveal the surplus generated by a trade. Subsequent work obtains positive results under smoothed valuation models~\cite{cesa2023repeated} or by aiming at $\alpha$-regret~\cite{azar2022alpha}.

Another line of work relaxes the requirement that the mechanism balance its budget in every round. \cite{bernasconi2024no} introduce global budget balance, which allows profits from some trades to subsidize others, and achieve $\widetilde{\mathcal{O}}(T^{3/4})$ regret with one-bit feedback even for adversarial valuations. \cite{chen2025tight} establish a matching lower bound and characterize the independent-valuation case through a tight $\widetilde{\Theta}(T^{2/3})$ rate. \cite{lunghi2025better} further characterize the tradeoff between regret and the allowed budget deficit. Crucially, the $T^{3/4}$ lower bound concerns GFT maximization and does not preclude faster rates for profit.

Related work extends the model along several directions. Contextual formulations allow valuations to depend on observable features through linear or Lipschitz functions~\cite{gaucher2025featurebased,coccianonparametric}; in the noiseless linear setting,~\cite{cosson2026contextual} study both GFT and profit maximization. Other extensions incorporate fairness in the division of surplus~\cite{bachoc2024fair}, introduce multiple buyers whose competition provides richer feedback~\cite{lunghi2025online}, or study the sample complexity of learning GFT-maximizing mechanisms in small markets~\cite{babaioff2024learning}.

\paragraph{The sample complexity of estimating multi-dimensional CDFs}

Assuming full feedback, classical bounds yield uniform CDF estimates using $\widetilde{\mathcal{O}}(\varepsilon^{-2})$ samples in any fixed dimension~\cite{dvoretzky1956asymptotic,massart1990tight,naaman2021tight}. Under one-bit feedback, the problem is closely related to learning demand curves through pricing queries: the learner posts a price and observes only whether a fresh buyer accepts it. Even in one dimension, finding an $\varepsilon$-optimal price for an arbitrary valuation distribution on $[0,1]$ requires $\widetilde{\Theta}(\varepsilon^{-3})$ queries~\cite{paseLeme}. \cite{castiglioni2026sample} study multidimensional CDF estimation under one-bit feedback. For an arbitrary joint distribution, they obtain a uniform $\varepsilon$-approximation over a grid with $K$ points per coordinate using $\varepsilon^{-3}\log(K/\delta)^{\mathcal{O}(d)}$ queries, with probability $1-\delta$. Thus, dimension affects only logarithmic factors, and the dependence on $\varepsilon$ matches the univariate lower bound up to such factors. Their result implies $\widetilde{\mathcal{O}}(T^{3/4})$ regret through an explore-then-commit reduction in any fixed dimension, leaving open whether this sample efficiency can translate into optimal regret guarantees.

\paragraph{Bandits with continuous action spaces}

Our problem is also related to continuum-armed bandits, where the learner optimizes an unknown reward function over a continuous domain. For Lipschitz rewards on $[0,1]^d$, the classical minimax regret rate is $\widetilde{\Theta}(T^{(d+1)/(d+2)})$, while adaptive discretization methods can obtain sharper guarantees when the near-optimal region has favorable geometry~\cite{bubeck2008online,kleinberg2019bandits}. In our setting, expected profit need not be Lipschitz, since the valuation distribution may have atoms. Nevertheless, monotonicity of the trade probability provides a one-sided discretization guarantee: rounding prices in the direction that facilitates trade loses only an amount proportional to the grid spacing. Combining this property with a standard bandit algorithm over a uniform grid recovers the $\widetilde{\mathcal{O}}(T^{3/4})$ baseline in two dimensions. Improving on this rate requires exploiting the additional structure of CDF-based rewards beyond this discretization argument.

\section{Preliminaries}\label{sec:model}

In this section, we formalize the learning problem and introduce the notation used throughout the paper.

\subsection{Learning in Bilateral Trade}

At each round $t\in[T]$, the intermediary posts a purchase price $p_t\in[0,1]$ to the seller and a sale price $q_t\in[0,1]$ to the buyer, without observing their private valuations. The valuation pairs $(S_t,B_t)$ are drawn independently across rounds from a common unknown joint distribution on $[0,1]^2$. Trade occurs when both agents accept their offers, namely, when $S_t\le p_t$ and $B_t\ge q_t$.

To highlight the connection with CDFs, we work on the transformed valuations and actions
\[
X_t=(S_t,1-B_t),
\qquad
x_t=(p_t,1-q_t),
\]
so that trade occurs exactly when $X_t\le x_t$, where vector inequalities are understood coordinatewise. Let $\mathcal L$ denote the common distribution of $X_t$, and let
\[
F(x)=\mathbb P_{X\sim\mathcal L}(X\le x)
\]
be its CDF. For any measurable set $A\subseteq[0,1]^2$, we use the shorthand
$
\mathbb P(A)=\mathbb P_{X\sim\mathcal L}(X\in A)
$. 

The intermediary receives only \emph{one-bit feedback}: it observes whether trade occurs, without observing either valuation or, when trade fails, which offer was rejected. Its observation is
\[
Y_t=\ind\{X_t\le x_t\}.
\]
For any action $x=(x_1,x_2)\in[0,1]^2$, the realized profit is
\[
r_t(x)=(1-x_1-x_2)\ind\{X_t\le x\}.
\]
For any action $x$, the expected profit is therefore
\[
r(x)=\E[r_t(x)]=(1-x_1-x_2)F(x).
\]

We benchmark the learner against the best fixed price pair, chosen with knowledge of $\mathcal L$ but without observing the realized valuations. The learner's \emph{expected cumulative regret} is
\begin{equation}\label{eq:regret-definition}
R_T
=
T\sup_{x\in[0,1]^2}r(x)
-
\E\left[\sum_{t=1}^T r(x_t)\right],
\end{equation}
where the expectation is over the random action sequence induced by the learner's internal randomization and past feedback.
The learner's goal is to minimize $R_T$.

\subsection{How to Discretize the Set of Prices }
\label{sec:discretization}

As is standard in bilateral trade~\cite{bernasconi2024no}, our algorithm operates on a finite discretization of the price space. We first introduce the discretization grid and the notation used throughout the paper, and then bound the approximation error incurred by restricting the learner to this grid.

For an integer $K\ge 2$, let $z_i=i/(K-1)$ for $i=0,\ldots,K-1$, and define the uniform grid
\[
\mathcal{Z}_K=\{z_0,\ldots,z_{K-1}\},
\qquad
\mathcal{X}_K=\mathcal{Z}_K\times\mathcal{Z}_K.
\]
For $i,j\in\{0,\ldots,K-1\}$, we write $x_{i,j}:=(z_i,z_j)$.  We also introduce the quantity
$
    \ell_K:=\log_2(K-1)+1,
$
which will be used in the subsequent analysis. For simplicity, we assume throughout that $\ell_K\in\mathbb{N}$, equivalently, that $K-1$ is a power of two. We will choose $K$ accordingly.

For $i,i',j\in\{0,\ldots,K-1\}$, we define the CDF difference between two points in the same row $j$ as
\[
    D_j(i,i'):=F(x_{i,j})-F(x_{i',j}).
\]
In the following, we will use estimates of these differences, denoted by $\widehat D_j(i,i')$.
%There are $K$ points along each coordinate, with spacing
%1/(K-1)$, and $K^2$ actions in total. As in the continuous
%model, a point $x=(z_i,z_j)$ corresponds to the price pair
%$(p,q)=(z_i,1-z_j)$.
%

We measure the suboptimality of a grid action relative to the best expected profit achievable on the grid. Specifically, we define
\[
\OPT=\max_{x\in\mathcal{X}_K}r(x),
\qquad
\gap(x)=\OPT-r(x)
\quad\text{for }x\in\mathcal{X}_K.
\]
Both quantities depend on the unknown distribution and
are used only in the analysis.

The following standard discretization lemma shows that restricting the learner to $\mathcal{X}_K$ incurs an approximation error of order $1/K$.

\begin{restatable}{lemma}{lemmaDisc}\label{lem:grid}
For every integer $K\ge2$, the optimal expected profit
on the grid satisfies the approximation bound
\[
0\le
\sup_{x\in[0,1]^2}r(x)-\OPT
\le\frac{2}{K-1}.
\]
\end{restatable}

Consequently, any algorithm selecting actions from $\mathcal{X}_K$ satisfies
\[
R_T
\le
\E\left[\sum_{t=1}^{T}\gap(x_t)\right]
+\frac{2T}{K-1}.
\]

\section{Main Theorem and Proof Roadmap}

In this paper, we will prove the following result. 

\begin{theorem}
\label{thm:main}
There exists an algorithm that achieves expected regret
\[
    R_T
    =
    \mathcal O\left(
    T^{2/3} \log^{9/2} T
    \right).
\]
\end{theorem}

The proof is organized as follows.

In Section~\ref{sec:successive-elimination}, we present our phased successive-elimination algorithm and analyze its regret. The algorithm repeatedly refines the CDF estimates of promising grid points and eliminates sufficiently suboptimal actions. We treat the refinement procedure as a black box, identifying the guarantees that suffice to establish the desired regret bound.

The remaining sections construct a refinement procedure satisfying these guarantees. In Section~\ref{sec:decomposing}, we introduce a mapped dyadic decomposition that allows us to reconstruct CDF differences between arbitrary points on the same row from a small collection of elementary CDF differences. Section~\ref{sec: char sample compl} develops a binary-search procedure for estimating these differences and characterizes its sample complexity in terms of the number of mapped-column changes and the cumulative probability mass of the associated regions.

In Section~\ref{sec:bent-columns}, we construct the mappings and establish the properties required to obtain a regret-efficient refinement procedure. In particular, we show that all necessary differences can be estimated using only near-optimal actions, while controlling both the number of representative replacements and the cumulative probability mass. We then combine these results to obtain accurate CDF estimates over the target set.

Finally, in Section~\ref{sec:refine}, we construct the complete
refinement procedure. We directly estimate the CDF
at one reference point per row and use the relative
estimates obtained in Section~\ref{sec:bent-columns} to reconstruct the
CDF values at all target points. We then show that
returning the target set with these refined estimates
preserves the conditions required by successive
elimination. This completes the construction of the procedure assumed in Section~\ref{sec:successive-elimination} and, consequently, the proof of Theorem~\ref{thm:main}.

\section{Phased Successive Elimination Algorithm}
\label{sec:successive-elimination}

In this section, we present our conservative phased successive-elimination algorithm and establish its regret guarantee. Starting from the discretization grid $\mathcal X_K$, the algorithm proceeds in phases, progressively refining the CDF estimates on an input set of grid actions. Each phase preserves all sufficiently near-optimal grid points, while ensuring that the actions queried during refinement have small suboptimality gaps. After the final phase, the algorithm commits to the action with the highest estimated expected profit.

We first analyze the outer successive-elimination algorithm, treating the refinement procedure as an oracle. This allows us to separate the regret analysis from the technical challenges of efficiently estimating the CDF. We specify the guarantees required from this oracle and show that they suffice to obtain the desired regret bound. 

\begin{definition}[Refinement oracle]
\label{def:refinement-oracle}
Fix $K\ge 2$, parameters $c_1,c_2>0$, and a positive
function $A=A(K,\delta)$.
A procedure $\textsc{Refine}$ is a
$(c_1,c_2,A)$-refinement oracle if, for every
$\varepsilon\in(0,1]$ and $\delta\in(0,1)$,
it takes as input $(\mathcal G,\widehat F,\varepsilon,\delta)$
satisfying
\[
\begin{aligned}
\{x\in\mathcal X_K:\gap(x)\le c_1\varepsilon\}
&\subseteq\mathcal G\subseteq\mathcal X_K,\\
\max_{x\in\mathcal G}|\widehat F(x)-F(x)|
&\le\varepsilon,
\end{aligned}
\]
and, with probability at least $1-\delta$, returns
$(\mathcal G',\widehat F')$ satisfying the same conditions
with precision $\varepsilon'=\varepsilon/8$.
On the same event, the procedure uses at most
\begin{equation}
\label{eq:refine-sample-complexity}
A(K,\delta)\left(
\varepsilon'^{-3}+K\varepsilon'^{-2}
\right)
\end{equation}
observations, each at an action $x$ satisfying
$\gap(x)\le c_2\varepsilon$.
\end{definition} 

The refinement oracle captures the properties required for successive elimination. Each call improves the precision of the CDF estimates while preserving all grid points that are sufficiently close to optimality at the next precision level. At the same time, the guarantees on sample complexity and the suboptimality gaps of queried actions control the regret incurred during exploration. As we show below, if $c_1,c_2$ are polylogarithmic in $K$ and $A(K,\delta)$ is polylogarithmic in $K/\delta$, these properties suffice to achieve $\widetilde{\mathcal{O}}(T^{2/3})$ expected regret.

For a horizon $T\geq2$, we choose
\begin{equation}
\label{eq:complete-parameters}
\begin{aligned}
K
=
2^{\lceil\log_2(T^{1/3})\rceil}+1, \quad
H
=
1+\lceil\log_8 T\rceil,\quad
\delta
=
\frac{T^{-3}}{H},\quad
\varepsilon_\star
=
4\sqrt{A(K,\delta)}\,T^{-1/3}.
\end{aligned}
\end{equation}

 Starting from the entire discretization grid, the algorithm repeatedly invokes the refinement oracle at geometrically decreasing precision levels. 
Each phase updates the input set and its CDF estimates, which are then used as input to the next phase. Refinement phases are executed until completion. If all phases are completed, the algorithm selects the action with the highest estimated expected profit and plays it for the remaining rounds.
The complete procedure is presented in Algorithm~\ref{alg:complete-algorithm}, whose guarantees are stated in the following theorem.

\begin{algorithm}[t]
\small
\caption{Phased successive elimination}
\label{alg:complete-algorithm}
\begin{algorithmic}[1]
\Require Horizon $T\geq2$

\State Compute $K,\delta,\varepsilon_\star$ using
\eqref{eq:complete-parameters}

\If{$\varepsilon_\star\geq1/8$}
    \State Play $x=(1,0)$ for all $T$ rounds
    \State \Return
\EndIf

\State Set
$m\gets\lfloor\log_8(1/\varepsilon_\star)\rfloor$

\State Set
$\mathcal G_0\gets\mathcal X_K$,
$\widehat F_0\equiv0$, and
$\varepsilon_0\gets1$

\For{$i=0,\ldots,m-1$}
    \State Run
    $\Call{Refine}{
        \mathcal G_i,\widehat F_i,
        \varepsilon_i,\delta}$
    \State \hspace{\algorithmicindent}
    \textit{Stop if the horizon is exhausted} 

    \State Let
    $(\mathcal G_{i+1},\widehat F_{i+1})$
    be the returned set and estimates

    \State Set
    $\varepsilon_{i+1}\gets\varepsilon_i/8$
\EndFor

\State Choose
\[
    x^\star\in
    \operatorname*{arg\,max}_{x\in\mathcal G_m}
    (1-x_1-x_2)\widehat F_m(x)
\]

\State Play $x^\star$ for all remaining rounds
\end{algorithmic}
\end{algorithm}

\begin{theorem}
\label{thm:main-parameters}
Suppose that $\textsc{\textnormal Refine}$ is a
$(c_1,c_2,A)$-refinement oracle, as defined in
Definition~\ref{def:refinement-oracle}. Then, with the parameters chosen as in \eqref{eq:complete-parameters}, Algorithm~\ref{alg:complete-algorithm}
achieves expected regret
\[
    R_T = \mathcal{O}\left(
        c_2\sqrt{A(K,\delta)}\,T^{2/3}
    \right).
\]

\end{theorem}

%\begin{theorem}
%\label{thm:main parameters}
%Under Assumption~\ref{assumption 1},
%Algorithm~\ref{alg:complete-algorithm} achieves expected regret
%\[
%    R_T
%    =
%    \mathcal O\left(
%    c_2\sqrt A\,T^{2/3}
%    \right).
%\]
%\end{theorem}

\begin{proof}
We first consider the case $\varepsilon_\star<1/8$.

Let $\mathcal E$ denote the event that all $m$ calls to \textsc{Refine} satisfy the guarantees of Definition~\ref{def:refinement-oracle}
in a hypothetical execution without a horizon restriction. The actual algorithm follows this execution until either all refinement phases are completed or the horizon is exhausted.

Since the initial set and estimates satisfy the input conditions of  Definition~\ref{def:refinement-oracle}, and each successful refinement guarantees the input conditions of the following phase, a conditional union bound gives
\[
    \mathbb P(\mathcal E)\geq1-m\delta.
\]

We bound the cumulative regret relative to the best grid action, defined as
\[
    Z_T:=\sum_{t=1}^{T}\gap(x_t).
\]
By Lemma~\ref{lem:grid}, the expected regret relative to the optimal continuous action satisfies
\[
    R_T
    \leq
    \mathbb E[Z_T]
    +\frac{2T}{K-1}.
\]

We first bound $Z_T$ on $\mathcal E$, distinguishing between exploration and commitment.

\paragraph{Exploration.}

Consider refinement phase $i$, for $i=1,\ldots,m$. On $\mathcal E$, this phase employs at most
\[
    A(K,\delta) \left(
    \varepsilon_i^{-3}
    +K\varepsilon_i^{-2}
    \right)
\]
observations, each at a point whose suboptimality gap is at most
\(
    c_2\varepsilon_{i-1}
    =
    8c_2\varepsilon_i.\)
Consequently, the total exploration regret is bounded by
\begin{align}
    Z_{\mathrm{explore}}
    \leq
    8c_2 A(K,\delta)
    \sum_{i=1}^{m}
    \left(
    \varepsilon_i^{-2}
    +K\varepsilon_i^{-1}
    \right)
    \leq
    16c_2A(K,\delta)
    \left(
    \varepsilon_m^{-2}
    +K\varepsilon_m^{-1}
    \right),
    \label{eq:exploration-regret-bound}
\end{align}
where the second inequality follows from the geometric decrease of the precision parameters. The same bound applies if the horizon expires during exploration, since the actual algorithm executes only a prefix of the hypothetical sequence of queries.

\paragraph{Commitment.}

Suppose that all refinement phases are completed before the horizon is exhausted.

On $\mathcal E$, the final set $\mathcal G_m$ contains a grid-optimal action, and its CDF estimates satisfy
\[
    \max_{x\in\mathcal G_m}
    |\widehat F_m(x)-F(x)|
    \leq\varepsilon_m.
\]

Since $|1-x_1-x_2|\leq1$, the corresponding estimated expected profits have estimation error at most $\varepsilon_m$.
Therefore, the action $x^\star$ maximizing the estimated profit over $\mathcal G_m$ satisfies
\[
    \gap(x^\star)\leq2\varepsilon_m.
\]

Its cumulative regret during commitment is thus bounded by
\(
    Z_{\mathrm{commit}}
    \leq2T\varepsilon_m.
\)
If instead the horizon expires during exploration, the algorithm does not enter the commitment stage, and hence $Z_{\mathrm{commit}}=0$.
Combining the exploration and commitment bounds, we conclude that, in either case, on $\mathcal E$,
\begin{equation}
\label{eq:successful-grid-regret}
    Z_T
    \leq
    16c_2 A(K,\delta)
    \left(
    \varepsilon_m^{-2}
    +K\varepsilon_m^{-1}
    \right)
    +2T\varepsilon_m.
\end{equation}

\paragraph{The complementary event.}

On $\mathcal E^c$, some refinement procedure may require more observations than its sample-complexity bound or return inaccurate estimates.

Nevertheless, the actual algorithm never executes more than $T$ rounds. Moreover, since $r(x)\in[-1,1]$, every action has suboptimality gap at most $2$. Thus,
\(Z_T\leq2T\) on every execution.

\paragraph{Bounding $Z_T$.} Using $\mathbb P(\mathcal E^c)\leq m\delta$ and \eqref{eq:successful-grid-regret}, we obtain
\begin{align*}
    \mathbb E[Z_T]
    &=
    \mathbb E[Z_T\mid\mathcal E]
    \mathbb P(\mathcal E)+
    \mathbb E[Z_T\mid\mathcal E^c]
    \mathbb P(\mathcal E^c)\leq
    16c_2A(K,\delta)
    \left(
    \varepsilon_m^{-2}
    +K\varepsilon_m^{-1}
    \right)
    +2T\varepsilon_m
    +2Tm\delta.
\end{align*}

Adding the discretization error gives
\begin{equation}
\label{eq:regret-before-parameters}
\begin{aligned}
    R_T
    \leq 
    16c_2A(K,\delta)
    \left(
    \varepsilon_m^{-2}
    +K\varepsilon_m^{-1}
    \right) +2T\varepsilon_m
    +2Tm\delta
    +\frac{2T}{K-1}.
\end{aligned}
\end{equation}

\paragraph{Setting the parameters.}

We choose $K$, $H$, $\delta$, and $\varepsilon_\star$ as in \eqref{eq:complete-parameters}. In particular,
\[
    K=\mathcal O(T^{1/3}),
    \qquad
    \delta=\mathcal O(T^{-3}),
    \qquad
    \varepsilon_\star=4\sqrt{A(K,\delta)}\,T^{-1/3}.
\]

Substituting these bounds into
\eqref{eq:regret-before-parameters} yields
\(
    R_T
    =
    \mathcal O\left(
    c_2\sqrt {A(K,\delta)}\,T^{2/3}
    \right).
\)

\paragraph{The case $\varepsilon_\star\geq1/8$.}

In this case, the algorithm plays the zero-profit action $x=(1,0)$ throughout the horizon. Since the optimal expected profit is at most one,
\(
    R_T\leq T.
\)

On the other hand,
$\varepsilon_\star\geq1/8$ implies
\[
    T
    \leq32\sqrt{A(K,\delta)}\,T^{2/3}.
\]
Thus, the same asymptotic regret bound holds in this case as well.
\end{proof}

To complete the proof of Theorem~\ref{thm:main} with regret $\mathcal O(T^{2/3}\log^{9/2}T)$, it remains to design a refinement oracle for parameters satisfying
\[
    c_1=\mathcal O(\log K),
    \qquad
    c_2=\mathcal O(\log K),
    \qquad
    A(K,\delta)=\mathcal O(\log^6(K) \log(K/\delta)).
\]

The remainder of the paper is devoted to constructing a refinement procedure satisfying these guarantees.

\section{Decomposing the CDF Differences} \label{sec:decomposing}

We begin the construction of the refinement procedure by showing how CDF differences between arbitrary points on the same row can be expressed as sums of a logarithmic number of elementary CDF differences. The key idea is to use a dyadic decomposition that allows different comparisons to share the same elementary CDF differences, thereby reducing the number of quantities that need to be estimated.

\subsection{Dyadic Decomposition} \label{sec:dyadic}
Following~\cite{castiglioni2026sample,CLM26b}, we introduce a family of dyadic pairs that can be used to decompose arbitrary intervals of grid indices.
Recall that $\ell_K=\log_2(K-1)+1$. We define the family
$\mathcal{I}^\star\subseteq
\{0,\ldots,K-1\}\times\{0,\ldots,K-1\}$
as
\begin{equation}
\label{eq:dyadic-family}
\mathcal{I}^\star :=
\bigcup_{k=0}^{\ell_K-2}
\left\{
(q2^k,(q+1)2^k):
q=0,\ldots,\frac{K-1}{2^k}-1
\right\}.
\end{equation}

The family contains pairs corresponding to consecutive dyadic intervals at different scales. By construction, its cardinality satisfies
\[
    |\mathcal{I}^\star|\leq 2K.
\]

The key property of this family is that any interval between two grid indices can be partitioned into a logarithmic number of dyadic intervals. Consequently, any CDF difference between points on the same row can be reconstructed by summing the differences associated with these intervals. We state this standard property in the following lemma and provide a proof in the appendix for completeness.

\begin{restatable}{lemma}{lemmaPartition} [Essentially Lemma 5.3 in \cite{CLM26b}]
\label{lem:dyadic-decomposition}
For every $i',i\in \{0,\ldots,K-1\}$ with $i'<i$, there exists a set
$\mathcal{B}(i,i')\subseteq\mathcal{I}^\star$, called the canonical dyadic decomposition and obtained by successively selecting the longest admissible dyadic interval, such that
\[
    \sum_{(a,b)\in\mathcal{B}(i,i')}(b-a)=i-i',
\]
and
\[
    |\mathcal{B}(i,i')|\leq 2\log_2(K-1).
\]
Moreover, the intervals corresponding to the pairs in \(\mathcal{B}(i,i')\) form a partition of \((i',i]\).
\end{restatable}

In particular, since the intervals in $\mathcal{B}(i,i')$
form a partition, for every row $j\in\{0,\ldots,K-1\}$ we have
\[
F(x_{i,j})-F(x_{i',j})
=
\sum_{(a,b)\in\mathcal{B}(i,i')}
\big(F(x_{b,j})-F(x_{a,j})\big).
\]
Thus, estimating the differences associated with the
$\mathcal{O}(K)$ pairs in $\mathcal{I}^\star$ is sufficient
to reconstruct any difference between points on the same row using at most
$\mathcal{O}(\log K)$ elementary CDF differences.

\begin{figure}
\centering
\begin{tikzpicture}[
  x=1cm, y=1cm,
  >=Stealth,
  font=\small,
  jump/.style={
    very thick, blue!70!black,
    -{Stealth[length=2.1mm]}
  },
  correction/.style={
    very thick, orange!85!black,
    -{Stealth[length=1.9mm]}
  },
  endpoint/.style={
    circle, fill=white, draw=blue!70!black,
    line width=1pt, inner sep=1.5pt
  }
]

\colorlet{mainblue}{blue!70!black}
\colorlet{accentorange}{orange!85!black}
\colorlet{sitecolor}{red!70!black}

% ================================================================
% Top panel
% ================================================================
\node[font=\normalsize\bfseries] at (6.5,8.4)
  {Dyadic endpoint jumps};

\draw[gray!55] (0,5.55) -- (13,5.55);
\foreach \k in {7,...,20} {
  \draw[gray!65] ({\k-7},5.43) -- ({\k-7},5.67);
  \node[below=4pt] at ({\k-7},5.43) {$\k$};
}

\foreach \k in {8,16,18,19} {
  \node[endpoint] at ({\k-7},5.55) {};
}

\draw[jump] (1,5.86)
  .. controls (2.4,7.40) and (7.6,7.40) .. (9,5.86);
\node[text=mainblue, fill=white, inner sep=1pt] at (5,7.25)
  {$8 \rightarrow 16$};

\draw[jump] (9,5.86)
  .. controls (9.32,6.75) and (10.68,6.75) .. (11,5.86);
\node[text=mainblue, fill=white, inner sep=1pt] at (10,6.92)
  {$16 \rightarrow 18$};

\draw[jump] (11,5.86)
  .. controls (11.18,6.45) and (11.82,6.45) .. (12,5.86);
\node[text=mainblue, fill=white, inner sep=1pt] at (11.5,6.63)
  {$18 \rightarrow 19$};

% ================================================================
% Bottom panel
% ================================================================
\node[font=\normalsize\bfseries] at (6.5,3.55)
  {Mapped representative index jumps};

\draw[gray!55] (0,0.55) -- (13,0.55);
\foreach \k in {7,...,20} {
  \draw[gray!65] ({\k-7},0.43) -- ({\k-7},0.67);
  \node[below=4pt] at ({\k-7},0.43) {$\k$};
}

% Occupied sites
\foreach \k in {7,9,10,11,12,13,14,15,17,19} {
  \draw[sitecolor, thick] ({\k-7},0.9) circle[radius=1.9pt];
}
\draw[sitecolor, thick] (1,0.9) circle[radius=1.9pt];

% Deleted sites
\foreach \k in {16,18} {
  \draw[sitecolor, thick]
    ({\k-7-.12},0.78) -- ({\k-7+.12},1.02)
    ({\k-7-.12},1.02) -- ({\k-7+.12},0.78);
}

% Initial correction
\draw[correction] (1,1.16)
  .. controls (.80,1.52) and (.20,1.52) .. (0,1.16);
\node[text=accentorange, anchor=south] at (1.5,1.3)
  {$8\rightarrow v_j(8)$};

% Main mapped jump
\draw[jump] (0,1.36)
  .. controls (1.2,2.90) and (6.8,2.90) .. (8,1.22);
\node[text=mainblue, fill=white, inner sep=1pt] at (4,3.05)
  {$v_j(8)\rightarrow v_j(16)$};

% Second mapped jump
\draw[jump] (8,1.18)
  .. controls (8.32,2.00) and (9.68,2.00) .. (10,1.18);
\node[text=mainblue, fill=white, inner sep=1pt] at (9,2.20)
  {$v_j(16)\rightarrow v_j(18)$};

% Third mapped jump
\draw[jump] (10,1.14)
  .. controls (10.20,1.62) and (11.80,1.62) .. (12,1.14);
\node[text=mainblue, fill=white, inner sep=1pt] at (11,1.82)
  {$v_j(18)\rightarrow v_j(19)$};

% Representative values: smaller labels avoid collisions
\foreach \k in {9,10,11,12,13,14,19,20} {
  \node[font=\scriptsize] at ({\k-7},-0.82) {$v_j(\k)$};
}

\node[font=\scriptsize] at (0,-0.58) {$v_j(7)$};
\node[font=\scriptsize] at (0,-1.05) {$v_j(8)$};

\node[font=\scriptsize] at (8,-0.58) {$v_j(15)$};
\node[font=\scriptsize] at (8,-1.05) {$v_j(16)$};

\node[font=\scriptsize] at (10,-0.58) {$v_j(17)$};
\node[font=\scriptsize] at (10,-1.05) {$v_j(18)$};

\end{tikzpicture}

\caption{
Dyadic decomposition and its mapped counterpart.
Blue arrows represent elementary CDF differences,
while the orange arrow represents the correction
introduced by the mapping. %\albo{Non ci andrebbe una X sull'8?}\anna{aggiunta frase per spiegare, non so se si capisce}
Notably, to maintain consistency across rows, the mapping may differ from the identity even when the original point is queryable, as illustrated by $v_j(8)$. Indeed, the mapping exhibits inertia across rows: once a mapped representative has been changed, it is not reset to its original position merely because that point becomes queryable again. 
}
\label{fig:dyadic-mapped-jumps}
\end{figure}
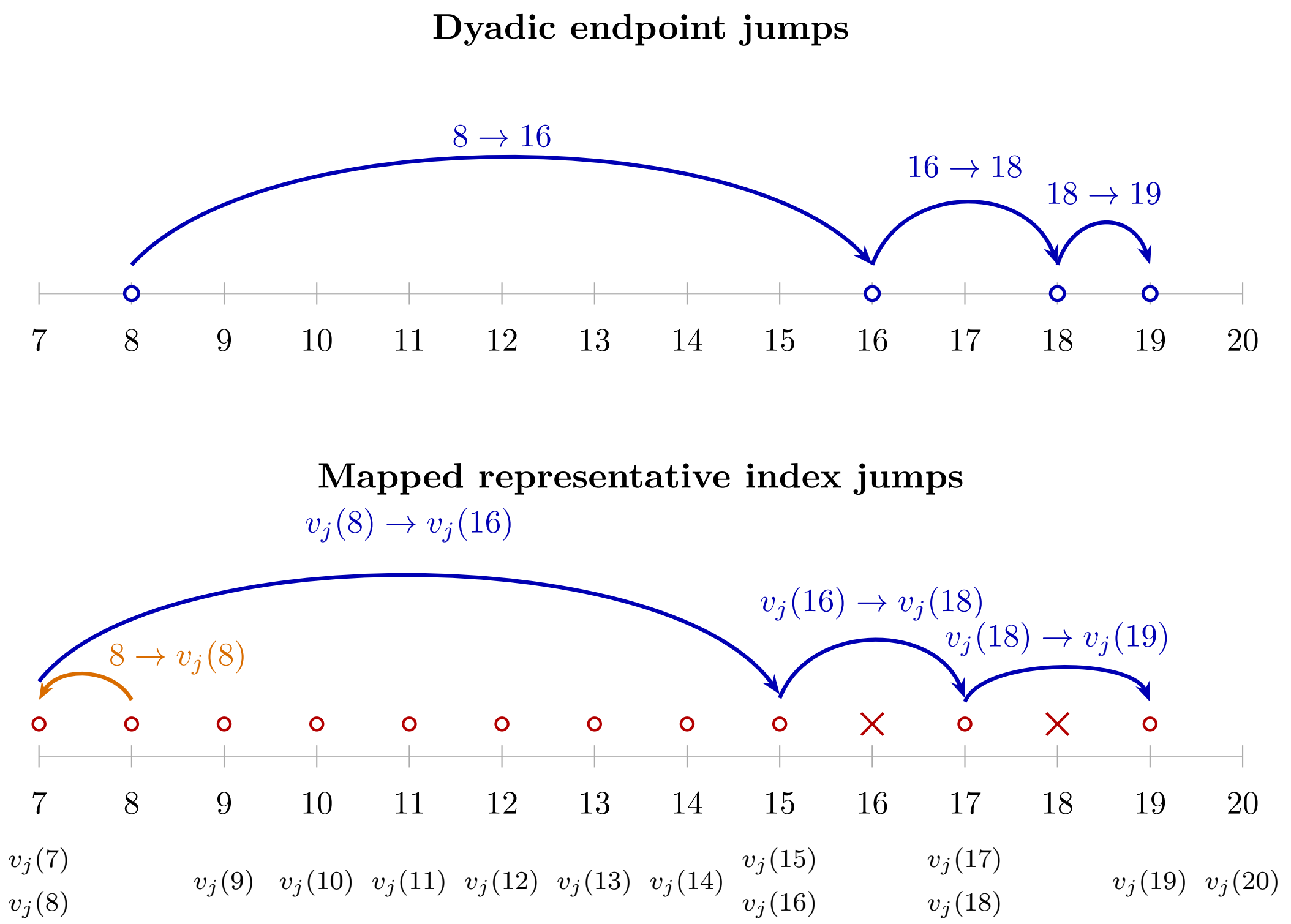

\subsection{Rerouting the Dyadic Decomposition}
\label{sec: rerouting map}
The dyadic decomposition introduced above allows us to reconstruct arbitrary same-row CDF differences from a small collection of elementary CDF differences. However, successive elimination may remove grid points whose CDF values are needed in this decomposition. To overcome this limitation, we introduce a modification that allows us to change the endpoints of the dyadic pairs while preserving the telescoping structure.

The key observation is that the decomposition relies only on the fact that the intervals in $\mathcal{B}(i,i')$ form a consecutive chain, rather than on the specific locations of their endpoints. Consequently, we can map the endpoints to different grid indices without affecting the validity of the decomposition.

Formally, for each row $j\in\{0,\ldots,K-1\}$, consider an arbitrary mapping
\[
    v_j:\{0,\ldots,K-1\}\rightarrow\{0,\ldots,K-1\}
\]
Then, for any two grid points $x_{i,j},x_{i',j}$ with $i'<i$, we have
\begin{align*}
    F(x_{i,j})-F(x_{i',j})
    =\bigl(F(x_{i,j})-F(x_{v_j(i),j})\bigr)&+\sum_{(a,b)\in\mathcal{B}(i,i')}
    \bigl(F(x_{v_j(b),j})-F(x_{v_j(a),j})\bigr)\\
    &+\bigl(F(x_{v_j(i'),j})-F(x_{i',j})\bigr).
\end{align*}
Indeed, the intermediate terms in the sum cancel because the pairs in $\mathcal{B}(i,i')$ form a consecutive chain, regardless of the choice of $v_j$.

This identity shows that, for each row $j$, all same-row CDF differences can be reconstructed from the collection of elementary CDF differences
\begin{align}\label{eq: base diff}
%\mathscr{D}_j :=
%&\left\{
F(x_{v_j(b),j})-F(x_{v_j(a),j}):
(a,b)\in\Dyad,
%\right\}\cup
%\left\{
\qquad
F(x_{i,j})-F(x_{v_j(i),j}):
i\in\{0,\ldots,K-1\}
%\right\}.
\end{align}

Since $|\Dyad|\leq 2K$, this collection contains at most $3K$ elementary CDF differences. Moreover, by the dyadic decomposition, any same-row CDF difference can be expressed as a sum of at most $2\log_2(K)+2$ elementary CDF differences.

Thus, introducing the mappings $v_j$ preserves both the linear number of elementary CDF differences and the logarithmic length of their decomposition, while providing additional flexibility in the choice of their endpoints. The challenge is to construct these mappings so that the required elementary CDF differences can be estimated efficiently using only queryable grid points.

Figure~\ref{fig:dyadic-mapped-jumps} illustrates the construction. The upper panel shows the original dyadic decomposition, while the lower panel shows how the endpoints are modified by the mapping $v_j$.

\section{Sample Complexity of Mapped Relative Learning}
\label{sec: char sample compl}

In this section, we isolate the properties that a family of mapped dyadic decompositions must satisfy in order to support efficient relative learning. This characterization explains the three requirements stated in the introduction and separates the sampling analysis from the construction of the mappings given in the next section.

Since
\[
F(x_{i,j})-F(x_{i',j})
=
\mathbb{P}\bigl(X\in(z_{i'},z_i]\times[0,z_j]\bigr),
\]
we can exploit binary search to efficiently estimate differences with fixed column endpoints across multiple rows. In fact, we can exploit the monotonicity of $\mathbb{P}\bigl(X\in(z_{i'},z_i]\times[0,z_j]\bigr)$ that can never decrease as $j$ increases.

We therefore first characterize the binary-search procedure over a consecutive block of rows in which the indices defining the elementary CDF differences remain unchanged, and prove that it achieves the desired estimation precision. We then characterize the total sample complexity of obtaining all the required estimates across all rows, accounting for the changes in the defining indices between consecutive blocks.

The resulting bound depends on two quantities: the number of fixed-endpoint blocks, controlled by the number of mapped-column changes, and the cumulative probability mass of their associated regions. The remaining requirement, queryability, guarantees that every elementary CDF difference needed to reconstruct a comparison between target points can be estimated using queryable actions when the corresponding block is sampled. Thus, the analysis below explains why the mapping construction must provide queryability, a small number of changes, and limited cumulative probability mass.

\subsection{Binary Search With Fixed Endpoints}

In this section, we focus on estimating the differences
\[
    \left\{
    F(x_{v_j(b),j})-F(x_{v_j(a),j})
    \right\}
\]
for a fixed pair $(a,b)\in\Dyad$ and for all
$j\in\{r_1,\ldots,r_2\}$, where
$r_1,r_2\in\{0,\ldots,K-1\}$ are such that
\[
    v_j(a)=v_{j'}(a), \qquad
    v_j(b)=v_{j'}(b),
    \qquad
    \forall j,j'\in\{r_1,\ldots,r_2\}.
\]
In other words, we consider a consecutive block of rows over which the mapped representative indices of the elementary CDF differences remain constant.

As we will discuss in more detail later, some of the points
$x_{v_j(a),j}$ and $x_{v_j(b),j}$ within this block may not belong to the queryable set. To understand why this may occur, consider the simple case in which an entire row $j$ contains no queryable actions. In this case, we are not interested in estimating the corresponding difference, since it will never be used to reconstruct a difference between two points in the target set.

Nevertheless, splitting the block at such a row would be unnecessary and potentially inefficient, as it would introduce additional binary-search initialization costs. A more efficient approach is to keep the entire block unchanged and simply mark row $j$ as ineligible for that block.

Our binary-search procedure must therefore accommodate blocks in which some of the mapped representatives are not queryable. More precisely, for each $j\in\{r_1,\ldots,r_2\}$, the procedure is required to return an estimate of
\[
    F(x_{v_j(b),j})-F(x_{v_j(a),j})
\]
only if both $x_{v_j(a),j}$ and $x_{v_j(b),j}$ belong to the queryable set.

The key property underlying this approach is that whenever a mapped representative $x_{v_j(i),j}$ is not queryable, no elementary CDF difference involving that point is needed to reconstruct the difference between any two points in the target set.

When we introduce the construction of the mappings $v_j(\cdot)$, we will prove that this property is indeed satisfied. Notice that this is precisely the queryability condition informally introduced in Requirement~\ref{req: 1}.

The same considerations apply to elementary CDF differences of the form
\[
    F(x_{v_j(i),j})-F(x_{i,j}),
\]
which appear as additional terms in the mapped decomposition.

Since the mapped representative indices remain constant throughout the block, we can denote them simply by $a,b\in\{0,\ldots,K-1\}$. In the following, we assume $a<b$. If the endpoints appear in the opposite order, we can exchange them and reverse the sign of the resulting estimates. If they coincide, the difference is identically zero and no estimation is required.

Let $\mathcal \Gqry\subseteq\mathcal{X}_K$ denote the queryable set, and let
\(
    \mathcal J=\{r_1,\ldots,r_2\}
\)
be the set of second-coordinate indexes corresponding to the rows of the block.

We aim to estimate the differences only at those rows for which both endpoints are queryable. We therefore define
\[
    W=
    \left\{
    w\in\mathcal J:
    (z_a,z_w),(z_b,z_w)\in\Gqry
    \right\}.
\]
If $W=\varnothing$, no estimate is required. Otherwise, let
\[
    w_0<w_1<\cdots<w_M
\]
denote the elements of $W$ in increasing order.

For convenience, define
\[
    P_{a,b}(w)
    :=F(x_{b,w})-F(x_{a,w})
    =\mathbb{P}\bigl(X\in(z_a,z_b]\times[0,z_w]\bigr).
\]
Notice that $P_{a,b}(w)$ is nondecreasing in $w$, even when some intermediate rows do not belong to $W$.

We now present a binary-search procedure that exploits this monotonicity to estimate $P_{a,b}(w)$ simultaneously for all $w\in W$, while collecting observations only at rows in $W$.

Then, we establish that Algorithm~\ref{alg:bins} achieves the desired estimation precision while sampling a number of rows that depends on the probability mass contained in the considered region, rather than on the number of eligible rows.

\begin{restatable}{lemma}{fixend}[Fixed-endpoint learning]
\label{lem:bins}
Fix $a<b$ in $\{0,\ldots,K-1\}$, a queryable set
$\Gqry\subseteq\mathcal{X}_K$, a nonempty set of eligible rows $W$, $\beta\in(0,1]$, and $\delta_0\in(0,1)$.
Algorithm~\ref{alg:bins} guarantees, with probability at least $1-\delta_0$, that
\[
    |\widehat P_{a,b}(w)-P_{a,b}(w)|\le\beta,
    \qquad \forall w\in W,
\]
and that the number of rows at which observations are collected is at most
\begin{equation}
\label{eq:bins-evaluations}
    2+\frac{2\lceil\log_2 K\rceil\mathbb{P}\bigl(X\in(z_a,z_b]\times(z_{\min W},z_{\max W}]\bigr)}{\beta}.
\end{equation}
Moreover, at each such row, the algorithm collects
\[
    n_B=
    \left\lceil32\beta^{-2}
    \log\frac{4K}{\delta_0}\right\rceil
\]
one-bit observations at each of the two endpoints.
\end{restatable}

\subsection{Aggregating Fixed-Endpoint Estimates}
\label{sec:pieces}

The previous lemma requires the two column coordinates to remain fixed throughout a call to \textsc{BinS}. Indeed, changing an endpoint changes the rectangle being measured, whose probability mass may decrease even as the row coordinate increases. We therefore partition the elementary CDF differences into consecutive blocks with fixed endpoints and apply \textsc{BinS} separately to each block, where the probability mass is nondecreasing with the row coordinate.

Let \(Q_j\) denote the set of queryable indices in row \(j\):
\[
Q_j := \{i \in \{0,\ldots,K-1\} : x_{i,j} \in \Gqry\}.
\]

Our goal is to estimate all the elementary CDF differences in

\begin{align*}
%\mathscr{D} = 
\bigcup_{j=0}^{K-1}
\Bigl(
&\left\{
D_j(v_j(b),v_j(a)):
(a,b)\in\Dyad,\;
v_j(a),v_j(b)\in Q_j
\right\}\cup
\left\{
D_j(i,v_j(i)):
i,v_j(i)\in Q_j
\right\}
\Bigr).
\end{align*}

\begin{comment}
\begin{align*}
\mathscr{D} =
&\left\{
F(x_{v_j(b),j})-F(x_{v_j(a),j}):
\begin{array}{l}
(a,b)\in\Dyad,\\
x_{v_j(a),j},x_{v_j(b),j}\in\Gqry,\\
j\in\{0,\ldots,K-1\}
\end{array}
\right\}\\
&\quad\cup
\left\{
F(x_{i,j})-F(x_{v_j(i),j}):
\begin{array}{l}
i,j\in\{0,\ldots,K-1\},\\
x_{i,j},x_{v_j(i),j}\in\Gqry
\end{array}
\right\}.
\end{align*}
\end{comment}
These are precisely the elementary CDF differences needed to reconstruct all same-row CDF differences between target points.

We call the row-dependent endpoint pair associated with each $(a,b)\in\Dyad$ a dyadic mapped column, and that associated with each index $i$ a correction mapped column. We partition these mapped columns into fixed-endpoint blocks, each consisting of a maximal consecutive set of rows over which the endpoints remain fixed.

More precisely, for each $(a,b)\in\Dyad$, we partition the rows into maximal consecutive blocks on which both $v_j(a)$ and $v_j(b)$ remain constant. Similarly, for each $i\in\{0,\ldots,K-1\}$, we partition the rows into maximal consecutive blocks on which $v_j(i)$ remains constant.

We denote by $\mathscr P$ the resulting collection of blocks. Each fixed-endpoint block $\rho\in\mathscr P$ is associated with two fixed column indexes $a_\rho,b_\rho\in\{0,\ldots,K-1\}$ and a consecutive set of rows. Within this set, we consider the differences
\[
    \left\{
    F(z_{b_\rho},z_j)-F(z_{a_\rho},z_j)
    \right\},
\]
restricted to the rows for which both endpoints belong to $\Gqry$.

As discussed in the previous section, ineligible rows do not split a block. Instead, we define $W_\rho$ as the set of second-coordinate values within the block for which both endpoints are queryable.
The advantage of this partition is that, for every block $\rho\in\mathscr P$ with $W_\rho\neq\varnothing$, we can apply the binary-search procedure introduced in the previous section.
Consequently, the total sample complexity depends on two quantities: the number of blocks, which determines the number of binary-search initializations, and the sum of the probability masses associated with the blocks, which determines the cost of the subsequent binary searches.

\paragraph{Number of blocks.}

The cardinality of $\mathscr P$ can be bounded directly in terms of the number of mapped-column changes.
Indeed, if all mappings $v_j(i)$ were constant across rows, we would have at most $|\Dyad|+K$ blocks: one for each dyadic mapped column and one for each correction mapped column.
Whenever a mapping changes between two consecutive rows, however, a new block may need to be initialized. We therefore define
\begin{align}
C :=
\sum_{j=0}^{K-2}
\Bigg(
&\sum_{(a,b)\in\Dyad}
\ind\left\{
(v_j(a),v_j(b))
\neq
(v_{j+1}(a),v_{j+1}(b))
\right\}
+
\sum_{i=0}^{K-1}
\ind\left\{
v_j(i)\neq v_{j+1}(i)
\right\}
\Bigg).
\label{eq:map-changes}
\end{align}

By construction, each change counted in $C$ introduces at most one additional block. Therefore,
\[
    |\mathscr P|
    \leq |\Dyad|+K+C
    \leq 3K+C.
\]

\paragraph{Cumulative probability mass.}

We now characterize the probability-dependent contribution to the sample complexity.

For each block $\rho\in\mathscr P$, we consider only the case $a_\rho<b_\rho$. If the original endpoints appear in the opposite order, we exchange them and negate the returned estimates to recover the desired signed differences. If the endpoints coincide, the corresponding difference is identically zero and no samples are required.

We therefore restrict attention to blocks with distinct endpoints and at least one eligible row.
For each such block $\rho$, let
\[
    w_1^\rho=\min W_\rho,
    \qquad
    w_2^\rho=\max W_\rho.
\]
We associate with $\rho$ the region
\begin{equation}
\label{eq:localized-mass}
    \mathsf R_\rho
    =
    (z_{a_\rho},z_{b_\rho}]
    \times
    (z_{w_1^\rho},z_{w_2^\rho}].
\end{equation}
We call $\mathbb P(X\in\mathsf R_\rho)$ its block mass and define the cumulative probability mass
\[
    \mathcal V
    :=
    \sum_{\rho\in\mathscr P}
    \mathbb{P}(X\in\mathsf R_\rho),
\]
where the sum ranges over blocks requiring a \textsc{BinS} call.
Notice that these regions may overlap. Consequently, $\mathcal V$ counts probability mass with multiplicity, and may be larger than one.

We can now express the sample complexity of estimating all the elementary CDF differences %in $\mathscr D$ 
in terms of $C$ and $\mathcal V$.

\begin{theorem}[Sampling fixed-endpoint blocks]
\label{thm:pieces}
Fix $\beta\in(0,1]$ and $\delta\in(0,1)$.

For each block $\rho\in\mathscr P$ with distinct endpoints and at least one eligible row, run Algorithm~\ref{alg:bins} with precision parameter $\beta$ and failure probability
\(
    \delta_0=\delta/({3K+C}).
\)
Then, with probability at least $1-\delta$, all returned estimates have estimation error at most $\beta$, simultaneously across all blocks and their eligible rows.
Moreover, on the same event, the total number of one-bit observations is at most
\begin{equation}
\label{eq:piece-cost}
    \mathcal O\left(
    \log\frac{4K(3K+C)}{\delta}
    \left[
        (3K+C)\beta^{-2}
        +
        \lceil\log_2 K\rceil
        \mathcal V\beta^{-3}
    \right]
    \right).
\end{equation}
\end{theorem}

\begin{proof}
By construction, the number of blocks requiring a call to \textsc{BinS} is at most
\[
    |\mathscr P|
    \leq |\Dyad|+K+C
    \leq 3K+C.
\]

For each such block $\rho$, the endpoints remain constant across the corresponding block of rows. Therefore, Lemma~\ref{lem:bins} applies with failure probability $\delta_0=\delta/({3K+C})$.

Applying a union bound over all calls, we obtain that, with probability at least $1-\delta$, all the returned estimates have estimation error at most $\beta$, and the bounds on the number of one-bit observations hold simultaneously.

Negating an estimate when the original endpoints are reversed does not change its estimation error. Moreover, differences with coinciding endpoints are identically zero and require no observations. Hence, the precision guarantee holds for all elementary CDF differences.

We now bound the total number of observations on this event.
By Lemma~\ref{lem:bins}, for each block $\rho$, the binary-search procedure collects observations at no more than
\[
    2+
    \frac{2\lceil\log_2 K\rceil}{\beta}
    \mathbb{P}(X\in\mathsf R_\rho)
\]
distinct rows.
Since there are at most $3K+C$ calls and
\[
    \mathcal V
    =
    \sum_{\rho\in\mathscr P}
    \mathbb{P}(X\in\mathsf R_\rho),
\]
the total number of rows at which observations are collected, counted separately for each call, is at most
\[
    2(3K+C)
    +
    \frac{2\lceil\log_2 K\rceil}{\beta}
    \mathcal V.
\]

At each sampled row, the algorithm collects an endpoint batch of $n_B$ one-bit observations at each of the two endpoints, where
\[
    n_B=
    \left\lceil
    32\beta^{-2}
    \log\frac{4K(3K+C)}{\delta}
    \right\rceil.
\]

Multiplying the bound on the number of sampled rows by $2n_B$, and using $\beta\leq1$ to absorb the ceiling into the constant factors, we obtain
\[
    \mathcal O\left(
    \log\frac{4K(3K+C)}{\delta}
    \left[
        (3K+C)\beta^{-2}
        +
        \lceil\log_2 K\rceil
        \mathcal V\beta^{-3}
    \right]
    \right),
\]
which proves \eqref{eq:piece-cost} and concludes the proof.
\end{proof}

\section{Constructing Mappings for Efficient Relative Learning}
\label{sec:bent-columns}

In this section, we construct mappings for a target set of grid points that meet the requirements characterized in Section~\ref{sec:overview}. Our approach combines the mapped dyadic decomposition introduced in Section~\ref{sec: rerouting map} with the fixed-endpoint binary-search procedure analyzed in Section~\ref{sec: char sample compl}.

The main idea is to construct the mappings $v_j(\cdot)$ so that every difference required to compare two target points on the same row can be reconstructed using only queryable points. %We then exploit these relative estimates to recover absolute CDF values by directly estimating one reference point for each row containing target points.

We first specify the inputs and the guarantees required by the construction. We then describe the mappings and prove that they satisfy the three requirements introduced in Section~\ref{sec:overview}: queryability, a small number of changes across rows, and limited cumulative probability mass. We conclude by applying the characterization of Section~\ref{sec: char sample compl} to obtain the relative-learning guarantee.

%\subsection{Inputs and Target Sets}

The refinement procedure is designed to operate over successive phases, each time receiving a subset of the original grid $\mathcal{X}_K$ containing all points whose suboptimality gaps are sufficiently small relative to the precision of the previous phase, together with uniformly accurate CDF estimates on this subset.

More precisely, at phase $i+1$, the procedure receives a set $\mathcal G\subseteq\mathcal{X}_K$ satisfying
\begin{equation}
\label{eq:refine-input-set}
    \left\{
    x\in\mathcal{X}_K:
    \operatorname{gap}(x)\leq128\varepsilon_i\ell_K
    \right\}
    \subseteq\mathcal G,
\end{equation}
together with estimates $\{\widehat F(x)\}_{x\in\mathcal G}$ such that
\begin{equation}
\label{eq:refine-input-estimates}
    \max_{x\in\mathcal G}
    |\widehat F(x)-F(x)|
    \leq\varepsilon_i.
\end{equation}
The goal is to return a new set and corresponding CDF estimates satisfying analogous guarantees at the improved precision
\[
    \varepsilon_{i+1}=\frac{\varepsilon_i}{8}.
\]

To achieve this efficiently, we distinguish two nested subsets of the input grid, based on the estimates obtained in the previous phase:

\begin{itemize}
    \item \textbf{Target set.}
    The first set, denoted by $\Gtar$, contains the points that may be sufficiently close to optimality to require retention in the next phase. In particular, it must contain every point whose suboptimality gap is at most
    \[
        128\varepsilon_{i+1}\ell_K
        =16\varepsilon_i\ell_K.
    \]
    Consequently, $\Gtar$ contains at least one optimal grid point. These are the points whose CDF estimates we aim to refine to precision $\varepsilon_{i+1}$.

    \item \textbf{Queryable set.}
    Estimating the CDF at the target points may benefit from querying additional points that are already known to be suboptimal. We therefore introduce a larger set, denoted by $\Gqry$, containing $\Gtar$ together with auxiliary points whose suboptimality gaps remain sufficiently small to make them affordable to query during the current phase. Although these additional points can be useful for estimating the CDF at the target points, they need not be retained in the next phase. Thus, we do not need to improve their CDF estimates.
\end{itemize}

The two sets satisfy
\[
\Gtar\subseteq\Gqry\subseteq\mathcal G.
\]
This distinction allows the procedure to exploit a larger set of queryable points while refining the estimates only on the smaller target set, thereby controlling both its sample complexity and the regret incurred during exploration.

Let
\[
    \widehat r(x)
    =(1-x_1-x_2)\widehat F(x),
    \qquad
    \widehat{\mathrm{OPT}}
    =\max_{x\in\mathcal G}\widehat r(x).
\]

We define the two subsets of $\mathcal G$ as follows:
\begin{equation}
\label{eq:refine-sets}
\begin{aligned}
\Gtar
&=
\left\{
x\in\mathcal G:
\widehat r(x)
\geq\widehat{\mathrm{OPT}}-20\varepsilon_i\ell_K
\right\},\\
\Gqry
&=
\left\{
x\in\mathcal G:
\widehat r(x)
\geq\widehat{\mathrm{OPT}}-100\varepsilon_i\ell_K
\right\}.
\end{aligned}
\end{equation}

The use of the queryable set $\Gqry$ introduces a separation between the target set $\Gtar$ and the unqueryable set $\mathcal{X}_K\setminus\Gqry$. The separation between these two sets is crucial. It allows us to change a mapped representative only when the previous mapped representative becomes sufficiently suboptimal, while ensuring that its replacement is significantly better.

We first record the consequences of the input conditions.

\begin{lemma}[Threshold separation]
\label{lem:threshold-separation}
Under conditions \eqref{eq:refine-input-set} and \eqref{eq:refine-input-estimates}, the following properties hold:
\begin{align*}
    \operatorname{gap}(x)
    &\leq22\varepsilon_i\ell_K,
    &&\forall x\in \Gtar,\\
    \operatorname{gap}(x)
    &\leq102\varepsilon_i\ell_K,
    &&\forall x\in \Gqry,\\
    \operatorname{gap}(x)
    &>98\varepsilon_i\ell_K,
    &&\forall x\notin \Gqry.
\end{align*}
Moreover, every point satisfying
\[
    \operatorname{gap}(x)\leq16\varepsilon_i\ell_K = 128\varepsilon_{i+1}\ell_K
\]
belongs to $\Gtar$.
\end{lemma}

\begin{proof}
Since $|1-x_1-x_2|\leq1$, the precision guarantee in
\eqref{eq:refine-input-estimates} implies
\[
    |\widehat r(x)-r(x)|\leq\varepsilon_i,
    \qquad \forall x\in\mathcal G.
\]

Furthermore, the input contains a grid-optimal point, so
\[
    |\widehat{\mathrm{OPT}}-\mathrm{OPT}|
    \leq\varepsilon_i.
\]
Consequently, estimated and true suboptimality gaps differ by at most $2\varepsilon_i$ on $\mathcal G$.

The first two claims follow immediately from the thresholds in \eqref{eq:refine-sets}, using $\ell_K\geq1$.
For a point $x\in\mathcal G\setminus \Gqry$, its estimated suboptimality gap exceeds $100\varepsilon_i\ell_K$, implying
\[
    \operatorname{gap}(x)
    >100\varepsilon_i\ell_K-2\varepsilon_i
    \geq98\varepsilon_i\ell_K.
\]
If instead $x\notin\mathcal G$, the input condition gives
$\operatorname{gap}(x)>128\varepsilon_i\ell_K$, which is stronger.

Finally, any point with gap at most $16\varepsilon_i\ell_K$ belongs to $\mathcal G$ by the input condition, and its estimated gap is at most
\[
    16\varepsilon_i\ell_K+2\varepsilon_i
    \leq18\varepsilon_i\ell_K
    <20\varepsilon_i\ell_K.
\]
It therefore belongs to $\Gtar$.
\end{proof}

\subsection{Mapping Construction}
\label{sec:map-construction}

We now describe how to construct the mappings $v_j(\cdot)$ used in the decomposition of Section~\ref{sec:decomposing}.
The construction follows three principles. First, the mappings are built sequentially across rows and remain unchanged whenever possible. Second, each mapped representative index is constrained to remain within a fixed neighborhood of its original position, determined by the dyadic structure. Third, a mapped representative is replaced only when it becomes unqueryable and a suitable replacement exists in the target set.

\paragraph{Permitted intervals.}

\begin{figure}[h]
\centering
\begin{tikzpicture}[
  x=.58cm,y=.78cm,>=Stealth,font=\small,
  support/.style={draw=mainblue!75!black,fill=mainblue!8,
                  line width=.75pt},
  junction/.style={circle,fill=sitecolor,draw=white,
                   line width=.4pt,inner sep=1.8pt},
  guide/.style={gray!22,line width=.3pt},
  level/.style={anchor=east,font=\footnotesize}
]
\colorlet{mainblue}{blue!70!black}
\colorlet{sitecolor}{red!70!black}

% Grid indices
\foreach \i in {0,...,16} {
  \draw[guide] (\i,-.35)--(\i,4.35);
  \fill[gray!60] (\i,-.15) circle (.8pt);
  \node[below=3pt,font=\scriptsize] at (\i,-.15) {$\i$};
}
\draw[gray!55] (0,-.15)--(16,-.15);

% h_i = 8
\node[level] at (-.55,3.75) {$h_i=8$};
\draw[support] (0,3.55) rectangle (16,3.95);
\node[junction] at (8,3.75) {};
\node[above=4pt,font=\footnotesize,black] at (8,3.8) {$I_8$};

% h_i = 4
\node[level] at (-.55,2.65) {$h_i=4$};
\foreach \a/\b/\i in {0/8/4,8/16/12} {
  \draw[support] (\a,2.45) rectangle (\b,2.85);
  \node[junction] at (\i,2.65) {};
  \node[above=4pt,font=\footnotesize,black] at (\i,2.65) {$I_{\i}$};
}

% h_i = 2
\node[level] at (-.55,1.55) {$h_i=2$};
\foreach \a/\b/\i in {0/4/2,4/8/6,8/12/10,12/16/14} {
  \draw[support] (\a,1.35) rectangle (\b,1.75);
  \node[junction] at (\i,1.55) {};
  \node[above=4pt,font=\footnotesize,black] at (\i,1.55) {$I_{\i}$};
}

% h_i = 1
\node[level] at (-.55,.45) {$h_i=1$};
\foreach \a/\b/\i in {0/2/1,2/4/3,4/6/5,6/8/7,
                       8/10/9,10/12/11,12/14/13,14/16/15} {
  \draw[support] (\a,.25) rectangle (\b,.65);
  \node[junction] at (\i,.45) {};
  \node[above=4pt,font=\scriptsize,black] at (\i,.5) {$I_{\i}$};
}

\node[anchor=west,font=\footnotesize,align=left] at (16.45,2.2)
  {$I_i=\{i-h_i,\ldots,i+h_i\}$,\\
   \quad $h_i=2^{\nu_2(i)}$.};

\end{tikzpicture}
\caption{Permitted intervals, grouped by their permitted radius.
Each red point is the junction \(i\) at the midpoint of its interval \(I_i\).}
\label{fig:dyadic-support-intervals}
\end{figure}
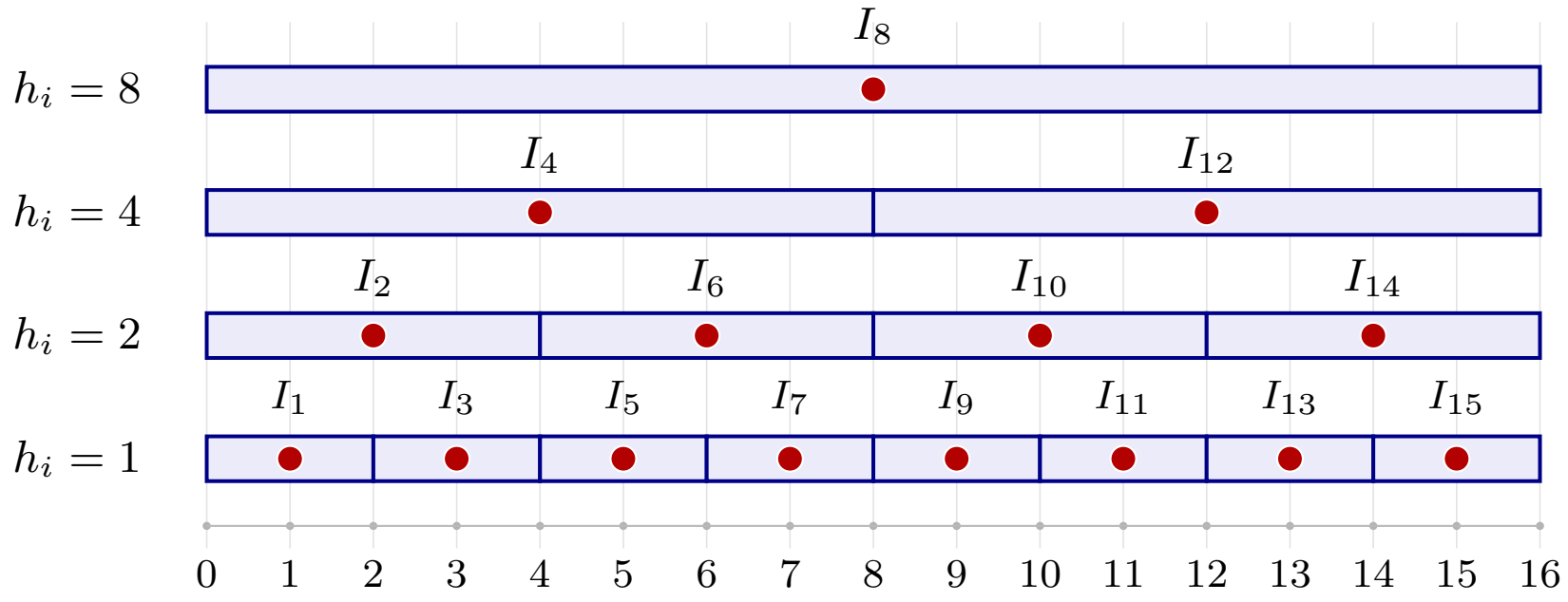

For every internal grid index $i\in\{1,\ldots,K-2\}$, let $\nu_2(i)$ denote the largest integer such that $2^{\nu_2(i)}$ divides $i$. Define
\[
    h_i:=2^{\nu_2(i)}
\]
as its permitted radius, and define the corresponding permitted interval
\begin{equation}
\label{eq:dyadic-support}
    I_i:=\{i-h_i,\ldots,i+h_i\}.
\end{equation}

The dyadic junction $i$ is the midpoint of a dyadic interval of length $2h_i$ (see Figure~\ref{fig:dyadic-support-intervals}). The set $I_i$ is the permitted interval within which its mapped representative index may move. The corresponding half-open coordinate interval $(z_{i-h_i},z_{i+h_i}]$ is its physical support.

For the boundary indices $0$ and $K-1$, we fix
\[
    v_j(0)=0,
    \qquad
    v_j(K-1)=K-1
\]
for every row $j$.

\paragraph{The mapping rule.}

We construct the mappings sequentially, starting from row $j=0$ and proceeding in increasing order.

For each internal index $i$, we initialize its mapped representative index at $i$. When processing row $j$, we preserve the current mapped representative index whenever the corresponding point belongs to $\Gqry$.
If the current mapped representative is not queryable, we look for a replacement inside $I_i$. More precisely, if there exists a point
\[
    x_{u,j}\in \Gtar,
    \qquad u\in I_i,
\]
we select one such index $u$ as the new mapped representative index.
If no such point exists, we leave the mapping unchanged, even if its current mapped representative is not queryable.
This last choice is important. As we will prove, when the permitted interval contains no target point, the corresponding mapped representative index is never needed to reconstruct a difference between two points in $\Gtar$ on that row.

The construction is summarized in Algorithm~\ref{alg:build-target-maps}.
\begin{algorithm}[t]
\small
\caption{\textsc{BuildTargetMaps}$(\Gtar,\Gqry)$}
\label{alg:build-target-maps}
\begin{algorithmic}[1]
\Require $\Gtar\subseteq \Gqry
\subseteq\mathcal{X}_K$
\Ensure Mappings $\{v_j(\cdot)\}_{j=0}^{K-1}$

\State Initialize $q_i\gets i$ for all $i=0,\ldots,K-1$

\For{$j=0,\ldots,K-1$}
    \For{$i=1,\ldots,K-2$}
        \If{$x_{q_i,j}\notin \Gqry$}
            \If{there exists $u\in I_i$ with
                 $x_{u,j}\in \Gtar$}
                \State Choose one such $u$
                \State $q_i\gets u$
            \EndIf
        \EndIf
    \EndFor
    \State Set $v_j(i)\gets q_i$ for all
    $i=0,\ldots,K-1$
\EndFor

\State \Return $\{v_j(\cdot)\}_{j=0}^{K-1}$
\end{algorithmic}
\end{algorithm}
Notice that the construction depends only on the input sets and does not require any new observations. Furthermore, each mapped representative index remains in its original permitted interval throughout the entire procedure.

\subsection{Queryability of the Required Differences}
\label{sec:target-reconstruction}

We now show that the mappings constructed above allow us to reconstruct every same-row CDF difference between points in $\Gtar$, while querying only points in $\Gqry$.
The key property is that every intermediate junction of the canonical dyadic decomposition has a permitted interval containing at least one endpoint of the original comparison.

\begin{lemma}[Canonical dyadic junctions]
\label{lem:target-junctions}
Fix two indices $a<b$ in $\{0,\ldots,K-1\}$.
The decomposition $\mathcal{B}(b,a)$ contains at most $2\log_2(K)$ intervals. Moreover, for every internal junction $k$ appearing between two consecutive intervals, its permitted interval $I_k$ contains at least one of $a$ and $b$.
\end{lemma}

\begin{proof}
The cardinality bound follows from the dyadic decomposition lemma of Section~\ref{sec:decomposing}.

For the second claim, consider an internal junction $k$ of the canonical decomposition. By construction, $I_k$ is the permitted interval of the largest centered dyadic interval having $k$ as its midpoint.

If neither $a$ nor $b$ belonged to $I_k$, then this centered interval would lie strictly inside the original comparison interval $(a,b]$. The dyadic intervals adjacent to $k$ could therefore be covered by a larger admissible dyadic interval, contradicting the maximality of the canonical decomposition.
Hence $I_k$ must contain at least one of the original endpoints.
\end{proof}

We can now establish the queryability of all the mapped differences required by the relative-learning procedure.

\begin{lemma}[Queryability of reconstruction representatives]
\label{lem:target-availability}
Fix a row $j$ and two target points
\[
    x_{a,j},x_{b,j}\in \Gtar,
    \qquad a<b.
\]

Every mapped representative appearing in the decomposition
\begin{align}
    F(x_{b,j})-F(x_{a,j})
    =
    \bigl(F(x_{b,j})-F(x_{v_j(b),j})\bigr)&+
    \sum_{(u,w)\in\mathcal B(b,a)}
    \bigl(
    F(x_{v_j(w),j})-F(x_{v_j(u),j})
    \bigr)
    \nonumber\\
    &+
    \bigl(F(x_{v_j(a),j})-F(x_{a,j})\bigr)
    \label{eq:target-mapped-decomposition}
\end{align}
belongs to $\Gqry$.
Consequently, every elementary CDF difference required by this decomposition can be estimated using only queryable points.
\end{lemma}

\begin{proof}
Consider an internal junction $i$ of the canonical dyadic decomposition.

By Lemma~\ref{lem:target-junctions}, its permitted interval $I_i$ contains at least one of the indices $a$ and $b$. Since both original endpoints belong to $\Gtar$, there exists at least one point of $\Gtar$ within the permitted interval of $i$ at row $j$.

By the mapping rule, either the previous mapped representative already belongs to $\Gqry$, in which case it is retained, or it is replaced by a point of $\Gtar\subseteq \Gqry$.
Thus,
\(
    x_{v_j(i),j}\in \Gqry.
\)

The same argument applies to the original comparison indices $a$ and $b$, since each belongs to its own permitted interval whenever it is internal. Boundary indices remain fixed, and their corresponding points belong to $\Gtar$ by assumption.
Therefore, all mapped representatives appearing in
\eqref{eq:target-mapped-decomposition} belong to $\Gqry$.
\end{proof}

The preceding lemma establishes Requirement~\ref{req: 1} for all the differences that our procedure actually needs to reconstruct.
In particular, a mapped representative may remain unqueryable on rows where its permitted interval contains no target point. This does not affect the reconstruction of any required difference, and allows the mappings to remain unchanged across otherwise unnecessary rows.

\subsection{Bounding the Number of Representative Replacements}
\label{sec:target-map-changes}

We now bound the number of representative replacements across consecutive rows.
The key observation is that every representative replacement corresponds to a sufficiently large improvement in expected profit. Since the profit function has bounded variation across the grid, such improvements cannot occur too frequently. For ease of notation let us define $v_{-1}$ as the identity map $v_{-1}(i)=i$ for all $i\in\{0,\ldots,K-1\}.$

\begin{lemma}[Gain from a representative replacement]
\label{lem:target-replacement-gain}
Assume that Equations \eqref{eq:refine-input-set} and \eqref{eq:refine-input-estimates} hold, and let  $v_0(\cdot),\ldots,v_{K-1}(\cdot)$ be the mappings constructed by Algorithm~\ref{alg:build-target-maps}.
Then, for every $i\in \{0,\ldots,K-1\},j\in\{0,\ldots,K-1\}$ such that $v_j(i)\neq v_{j-1}(i)$, it holds
\[
    r(x_{v_{j}(i),j})-r(x_{v_{j-1}(i),j})
    >76\varepsilon_i\ell_K.
\]
\end{lemma}

\begin{proof}
By the update rule, the new mapped representative belongs to $\Gtar$, whereas the previous mapped representative lies outside $\Gqry$.

Lemma~\ref{lem:threshold-separation} therefore gives
\(
    \operatorname{gap}(x_{v_j(i),j})
    \leq22\varepsilon_i\ell_K
\)
and
\(
    \operatorname{gap}(x_{v_{j-1}(i),j})
    >98\varepsilon_i\ell_K.
\)
Consequently,
\begin{align*}
    r(x_{v_j(i),j})-r(x_{v_{j-1}(i),j})>76\varepsilon_i\ell_K.
\end{align*}
\end{proof}

For any mapping $v$ satisfying
$v(i)\in I_i$, define the profit potential
\begin{equation}
\label{eq:potential-definition}
    \Phi(v,j)
    :=
    \sum_{i=1}^{K-2}
    \left(
        r(x_{v(i),j})-r(x_{i-h_i,j})
    \right).
\end{equation}

The potential depends on both the mapping and the row.
Our objective is to bound the cumulative increase in the
potential caused by representative replacements.

In particular, thanks to Lemma \ref{lem:target-replacement-gain} we know that 
\[\sum_{j=1}^{K-1}\left(\Phi(v_j(\cdot),j)-\Phi(v_{j-1}(\cdot),j)\right)\ge 76\varepsilon_i \ell_K\left(\sum_{j=1}^{K-1}\sum_i \ind\{v_j(i)\neq v_{j-1}(i)\}\right).\]

To bound the number of mapping changes, it remains to
control the cumulative increase in the potential induced
by these changes, namely,
\[\sum_{j=0}^{K-1}\left(\Phi(v_j(\cdot),j)-\Phi(v_{j-1}(\cdot),j)\right)\]
where we recall that $v_{-1}$ is the identity
mapping, i.e., $v_{-1}(i)=i$ for all
$i\in\{0,\ldots,K-1\}$.
To this end, we decompose the sum as follows:%\begin{equation}
\begin{align}
\label{eq: decomp phi}
    \sum_{j=0}^{K-1}\left(\Phi(v_j(\cdot),j)-\Phi(v_{j-1}(\cdot),j)\right)&= \underbrace{\Phi(v_{K-1}(\cdot),K-1)}_{\Phi_{\mathrm{final}}}-\underbrace{\Phi(v_{-1}(\cdot),0)}_{\Phi_{\mathrm{initial}}} \\
    & \quad- \sum_{j=0}^{K-2}\left(\Phi(v_{j}(\cdot),j+1)-\Phi(v_{j}(\cdot),j)\right)
\end{align}
%\end{equation}
and we bound each term separately.
To do so, we will use the following lemma. 
\begin{lemma}[Potential range and row variation]
\label{lem:potential-drift}
For every admissible mapping $v(\cdot)$ and every row $j$, we have
\[
    -2\ell_K\leq\Phi(v(\cdot),j)\leq\ell_K.
\]
Moreover, for every sequence of admissible mappings
$v_0(\cdot),\ldots,v_{K-1}(\cdot)$% and every $j=0,\ldots,K-2$,
%
%\label{eq:one-step-row-drift}
%\begin{align*}
%    |\Phi(v_j(\cdot),j+1)-\Phi(v_j(\cdot),j)|\leq
%    2\mathbb P\left(
%        X\in[0,1]\times (z_j,z_{j+1}]
%    \right)
%    +(z_{j+1}-z_j).
%\end{align*}
%Consequently,
\begin{equation}
\label{eq:cumulative-row-drift}
    \sum_{j=0}^{K-2}
    |\Phi(v_j(\cdot),j+1)-\Phi(v_j(\cdot),j)|
    \leq3 \ell_K.
\end{equation}
\end{lemma}

\begin{proof}
Let us fix a permitted radius $h$. If we consider only the internal dyadic junctions with permitted radius $h$, the associated intervals $(z_{i-h_i},z_{i+h_i}]$ are disjoint.

\textbf{Potential range.}
Fix an admissible mapping $v(\cdot)$ and a row $j$.
Using $r(x)=(1-x_1-x_2)F(x)$, we have
\begin{align*}
r(x_{v(i),j})-r(x_{i-h,j})
&=
(1-z_{v(i)}-z_j)
\left(
    F(x_{v(i),j})-F(x_{i-h,j})
\right)-
(z_{v(i)}-z_{i-h})F(x_{i-h,j}).
\end{align*}

Since $v(i)\in I_i$, the intervals $(z_{i-h},z_{v(i)}]$ for all $i$ such that $h_i=h$ are disjoint and contained in $[0,1]$.
Therefore,
\[
    \sum_{i:h_i=h}
    \left(
        F(x_{v(i),j})-F(x_{i-h,j})
    \right)
    \leq1
\]
and
\[
    \sum_{i:h_i=h}
    (z_{v(i)}-z_{i-h})
    \leq1.
\]

Since $|1-z_{v(i)}-z_j|\leq1$ and
$0\leq F(x_{i-h,j})\leq1$, the first
contribution to the potential belongs to
$[-1,1]$, while the second belongs to
$[-1,0]$.
Consequently, considering that there are $\ell_K$ possible values of $h$, 
\[
    -2\ell_K\leq\Phi(v,j)\leq \ell_K.
\]

\textbf{Variation between consecutive rows.}
Fix $j\in\{0,\ldots,K-2\}$.
We consider the variation
\[
    \Phi(v_j(\cdot),j+1)-\Phi(v_j(\cdot),j),
\]
where the mapping $v_j(\cdot)$ remains fixed.
For each $i$ such that $h_i=h$, we have
\begin{align*}
r(x_{v_j(i),j+1})-r(x_{i-h,j+1})-r(x_{v_j(i),j})&+r(x_{i-h,j})\\
&=
(1-z_{v_j(i)}-z_{j+1})
\mathbb P\left(
    X\in
    (z_{i-h},z_{v_j(i)}]
    \times(z_j,z_{j+1}]
\right)\\
&\quad-
(z_{j+1}-z_j)
\mathbb P\left(
    X\in
    (z_{i-h},z_{v_j(i)}]
    \times[0,z_j]
\right)\\
&\quad-
(z_{v_j(i)}-z_{i-h})
\mathbb P\left(
    X\in
    [0,z_{i-h}]
    \times(z_j,z_{j+1}]
\right).
\end{align*}

Summing over all $i$ with $h_i=h$, we bound
the absolute contributions of these three
terms separately.

For the first term, since
$|1-z_{v_j(i)}-z_{j+1}|\leq1$ and the intervals
$(z_{i-h},z_{v_j(i)}]$ are disjoint, its
absolute contribution is bounded by
\[
    \mathbb P\left(X\in[0,1]\times (z_j,z_{j+1}]\right).
\]

The contribution of the second term  can be bounded by $z_{j+1}-z_j$, since the same disjointness
implies
\begin{align*}
    &\sum_{i:h_i=h}
    \mathbb P\left(
        X\in
        (z_{i-h},z_{v_j(i)}]
        \times[0,z_j]
    \right)
    \leq1.
\end{align*}
%Its absolute contribution is therefore
%at most $z_{j+1}-z_j$.

Finally, for the third term, we use
\[
    \sum_{i:h_i=h}
    (z_{v_j(i)}-z_{i-h})
    \leq1
\]
and

\begin{align*}
    &\mathbb P\left(X\in[0,z_{i-h}]\times(z_j,z_{j+1}]\right)\le\mathbb P\left(X\in[0,1]\times (z_j,z_{j+1}]\right).
\end{align*}

Thus, the absolute contribution of the
third term is also at most $\mathbb P\left(X\in[0,1]\times (z_j,z_{j+1}]\right)$.
Combining these bounds gives
\begin{align*}
    |\Phi(v_j,j+1)-\Phi(v_j,j)|\leq
    \ell_K\left(2\mathbb P\left(
        X_2\in(z_j,z_{j+1}]
    \right)
    +(z_{j+1}-z_j)\right).
\end{align*}

\textbf{Cumulative row variation.}
Summing over all row transitions,
\begin{align*}
    &\sum_{j=0}^{K-2}
    |\Phi(v_j,j+1)-\Phi(v_j,j)|\leq
    \ell_K 2\sum_{j=0}^{K-2}
    \mathbb P\left(
        X_2\in(z_j,z_{j+1}]
    \right)+
    \ell_K\sum_{j=0}^{K-2}(z_{j+1}-z_j)\leq\ell_K(2+1)=3\ell_K.
\end{align*}

The first sum is at most one because
the intervals $(z_j,z_{j+1}]$ are disjoint,
while the second telescopes to one.
Notice that the mappings $v_j$ may change
between row transitions: the argument only
requires that each mapping remains inside
its prescribed permitted intervals.
\end{proof}

Thanks to Lemma \ref{lem:potential-drift} we can bound all three terms of \eqref{eq: decomp phi}, the sum of the first two by $\ell_K-(-2)\ell_K=3\ell_K$ and the third one by $3\ell_K$ thanks to \eqref{eq:cumulative-row-drift}.  

\begin{lemma}[Cumulative representative replacement gain]
\label{lem:cumulative-replacement-gain-2}
Let $v_{-1}(\cdot)$ denote the identity mapping used
to initialize the construction, and let
$v_0(\cdot),\ldots,v_{K-1}(\cdot)$ be the mappings obtained
after processing the corresponding rows.
Then
\begin{equation}
\label{eq:cumulative-replacement-gain}
    \sum_{j=0}^{K-1}
    \left(\Phi(v_j,j)-\Phi(v_{j-1},j)\right)
    \leq 6\ell_K.
\end{equation}
\end{lemma}

\begin{proof}
The quantity in \eqref{eq:cumulative-replacement-gain}
is the cumulative change in the potential
caused by updating the mappings, with the
row coordinate held fixed during each update.
By adding and subtracting the potential
evaluated at the previous mapping, we obtain
the telescoping identity of \eqref{eq: decomp phi}.
We bound the three terms separately.

By Lemma~\ref{lem:potential-drift},
\(
    \Phi_{\mathrm{final}}
    =
    \Phi(v_{K-1},K-1)
    \leq \ell_K,
\)
\(
    \Phi_{\mathrm{initial}}
    =
    \Phi(v_{-1}(\cdot),0)
    \geq-2\ell_K,
\)
and
\(
    \sum_{j=0}^{K-2}
    |\Phi(v_j,j+1)-\Phi(v_j,j)|
    \leq3\ell_K.
\)
Substituting these bounds into \eqref{eq: decomp phi}, we conclude the proof.
\end{proof}

Together with Lemma \ref{lem:cumulative-replacement-gain-2}, we can use Lemma \ref{lem:target-replacement-gain} to bound the number of representative replacements.

\begin{lemma}[Number of representative replacements]
\label{lem:number-of-replacements}
By construction, we have
\[
    \sum_{j=1}^{K-1}\sum_{i=0}^{K-1}\ind\{v_j(i)\neq v_{j-1}(i)\}
    \leq
    \frac{6}{76\varepsilon_i}.
\]
\end{lemma}

\begin{proof}
By Lemma~\ref{lem:target-replacement-gain}, each representative replacement changes
the mapped representative from a point outside
$\Gqry$ to a point in $\Gtar$, increasing its expected
profit by more than $76\varepsilon_i\ell_K$.
Since each representative replacement changes exactly one
summand of the potential, its contribution
to the cumulative representative replacement gain is
greater than $76\varepsilon_i\ell_K$.
Therefore,
\begin{align*}
    76\varepsilon_i\ell_K \left(\sum_{j=1}^{K-1}\sum_{i=0}^{K-1}\ind \{v_j(i)\neq v_{j-1}(i)\}\right)
    &\le
    \sum_{j=0}^{K-1}
    \left(
        \Phi(v_j(\cdot),j)-\Phi(v_{j-1}(\cdot),j)
    \right)\leq 6\ell_K,
\end{align*}
which yields
\[
    \sum_{j=1}^{K-1}\sum_{i=0}^{K-1}\ind\{v_j(i)\neq v_{j-1}(i)\}
    \leq
    \frac{6}{76\varepsilon_i}.
\]
\end{proof}

\paragraph{From representative replacements to fixed-endpoint blocks.}

Since each internal $v_j(i)$ appears as an endpoint of at most two dyadic intervals at each scale, a single representative replacement can affect at most $2(\ell_K-1)$ dyadic mapped columns and one correction mapped column.

Let $C$ denote the total number of mapped-column changes, as defined in Section~\ref{sec:pieces}.
We therefore obtain
\[
    C
    \leq(2\ell_K-1)\sum_{j=1}^{K-1}\sum_{i=0}^{K-1}\ind\{v_j(i)\neq v_{j-1}(i)\}
    =
    \mathcal O\left(\frac{\ell_K}{\varepsilon_i}\right).
\]

Since there are at most $|\mathcal I^\star|+K\leq3K$ initial mapped columns, the total number of fixed-endpoint blocks is bounded by
\begin{equation}
\label{eq:target-piece-count}
    |\mathscr P|
    \leq3K+C
    =
    \mathcal O\left(
    K+\frac{\ell_K}{\varepsilon_i}
    \right).
\end{equation}

This establishes Requirement~\ref{req: 2}.

\subsection{Bounding Cumulative Probability Mass}

It remains to bound the cumulative probability mass
$\mathcal V$ of the rectangular regions introduced in Section~\ref{sec: char sample compl}.

The main difficulty is that mapped representatives may change across rows, so the corresponding regions need not be
disjoint. Nevertheless, each mapped representative index remains within its fixed permitted interval. We exploit this locality to bound the number of times any probability mass can be counted.

By construction $v_j(i)\in \{i-h_i,\ldots,i+h_i\}$ for every row $j$ and consequently,
the horizontal interval associated with the
correction mapped column
\((i,v_j(i))\) is contained in $[z_{i-h_i},z_{i+h_i}]$, regardless of how the
mapping changes across rows.

We show that the same property holds for dyadic
mapped columns: each can be assigned to a fixed
interval containing its horizontal difference
interval at every row.

\begin{lemma}[Fixed interval]
\label{lem:target-enclosing-supports}
For every $i\in \{1,\ldots,K-2\}$ and every row $j\in \{0,\ldots,K-1\}$,
\[
\begin{aligned}
    (&\min\{z_i,z_{v_j(i)}\},\max\{z_i,z_{v_j(i)}\}]
    \subseteq (z_{i-h_i},z_{i+h_i}].
\end{aligned}
\]

Furthermore, every $(a,b)\in\mathcal I^\star$ can be
assigned to an interval of the type $(z_{i-h_i},z_{i+h_i}]$ whose closure
contains both $z_{v_j(a)}$ and $z_{v_j(b)}$ for
every row $j$. In particular,
\[
\begin{aligned}
    (&\min\{z_{v_j(a)},z_{v_j(b)}\},\max\{z_{v_j(a)},z_{v_j(b)}\}]
\end{aligned}
\]
is contained in the interval $(z_{i-h_i},z_{i+h_i}]$, where $i\in\{a,b\}$ is the endpoint with the larger permitted radius.
Each interval of the type $(z_{i-h_i},z_{i+h_i}]$ is assigned at most
$2(\ell_K-1)$ dyadic mapped columns and
one correction mapped column.
\end{lemma}

\begin{proof}
Consider first the correction mapped column associated
with an internal junction $i$.

By construction,
\[
    v_j(i)\in I_i
    =
    \{i-h_i,\ldots,i+h_i\}
\]
for every row $j$. Hence,
\[
    z_i,z_{v_j(i)}
    \in[z_{i-h_i},z_{i+h_i}].
\]
We now consider a dyadic mapped column
associated with a pair
$(a,b)\in\mathcal I^\star$ of length $2^k$.

Suppose first that both $a$ and $b$ are internal
indices. Since they are consecutive multiples
of $2^k$, exactly one is an odd multiple of
$2^k$ and therefore has permitted radius
$2^k$. The other is divisible by a larger power
of two and has a permitted interval containing
that of the former.

We assign the dyadic mapped column to the interval $(z_{i-h_i},z_{i+h_i}]$
where $i$ is the endpoint with the larger permitted
interval. Since both $v_j(a)\in I_a$ and
$v_j(b)\in I_b$, the mapped endpoint coordinates
$z_{v_j(a)}$ and $z_{v_j(b)}$ belong to the
closure of this interval $[z_{i-h_i},z_{i+h_i}]$ for every row $j$.
Consequently, their sorted half-open difference
interval is contained in the designated interval $(z_{i-h_i},z_{i+h_i}]$.

If one of $a,b$ is a boundary index, we assign
the mapped column to the interval $(z_{i-h_i},z_{i+h_i}]$ of the
internal endpoint. The corresponding permitted
interval contains the boundary, whose mapping
is fixed. The same containment property
therefore holds.

Finally, each internal junction is an endpoint
of at most two dyadic intervals at each scale.
Since $\mathcal I^\star$ contains
$\ell_K-1$ scales, at most $2(\ell_K-1)$
dyadic mapped columns can be assigned
to the same $(z_{i-h_i},z_{i+h_i}]$ of any internal junction $i$.

Adding its correction mapped column completes
the proof.
\end{proof}

This allows us to control
the cumulative probability mass even though
regions associated with different mapped columns may
overlap.

\begin{theorem}[Cumulative probability mass]
\label{thm:target-overlap}
Let $\mathscr P$ denote the collection of
blocks induced by the mappings
constructed in Algorithm~\ref{alg:build-target-maps}.
For each block $\rho$, let $\mathsf R_\rho$
be its associated probability region.
Then,
\[
    \mathcal V \coloneqq \sum_{\rho\in\mathscr P}
    \mathbb P(X\in\mathsf R_\rho)
    \leq
    (2\ell_K-1)(\ell_K-1)
    =
    \mathcal O(\ell_K^2).
\]
%Moreover, every point $x\in[0,1]^2$ belongs
%to at most $(2\ell_K-1)(\ell_K-1)$
%of the regions $\mathsf R_\rho$.
\end{theorem}

\begin{proof}
We first bound the number of regions containing
an arbitrary point $x\in[0,1]^2$.

Each block is obtained by partitioning the rows into maximal consecutive
blocks on which the endpoints of the elementary CDF differences remain unchanged.

For each $(a,b)\in\mathcal I^\star$, consider the
blocks associated with the differences
\[
    F(x_{v_j(b),j})-F(x_{v_j(a),j}).
\]
Since the corresponding fixed-endpoint blocks are disjoint,
the vertical intervals of their associated regions
$\mathsf R_\rho$ are also disjoint.
Consequently, any point $x\in[0,1]^2$ can belong
to at most one such region for each fixed pair
$(a,b)$.

The same argument applies to the blocks associated
with the differences
\[
    F(x_{v_j(i),j})-F(x_{i,j})
\]
for each fixed internal index $i$.

By Lemma~\ref{lem:target-enclosing-supports},
all the regions associated with a fixed pair
$(a,b)\in\mathcal I^\star$ are horizontally
contained in the same interval $(z_{i-h_i},z_{i+h_i}]$.
The same holds for the regions associated
with each fixed internal index $i$.
Moreover, each interval $(z_{i-h_i},z_{i+h_i}]$ is associated
with at most $2(\ell_K-1)$ pairs from
$\mathcal I^\star$ and one internal index.
Therefore, any point $x$ can belong to at most
\[
    2(\ell_K-1)+1=2\ell_K-1
\]
regions associated with a given interval of the type $(z_{i-h_i},z_{i+h_i}]$.
Finally, the intervals $(z_{i-h_i},z_{i+h_i}]$ at any fixed
dyadic scale $h_i=h$ are disjoint, and there are
$\ell_K-1$ scales.
Hence, the first coordinate $x_1$ belongs
to at most one $(z_{i-h_i},z_{i+h_i}]$ per scale.

Combining these observations, we obtain
\begin{equation}
\label{eq:pointwise-overlap}
    \sum_{\rho\in\mathscr P}
    \ind\{x\in\mathsf R_\rho\}
    \leq
    (2\ell_K-1)(\ell_K-1)
\end{equation}
for every $x\in[0,1]^2$.
Taking expectations with respect to
$X$ on both sides of
\eqref{eq:pointwise-overlap} yields
\begin{align*}
    \mathcal V=
    \sum_{\rho\in\mathscr P}
    \mathbb P(X\in\mathsf R_\rho)=
    \mathbb E\left[
        \sum_{\rho\in\mathscr P}
        \ind\{X\in\mathsf R_\rho\}
    \right]\leq
    (2\ell_K-1)(\ell_K-1)=
    \mathcal O(\ell_K^2).
\end{align*}
%This proves the claim.
\end{proof}

This establishes Requirement~\ref{req: 3}.

\subsection{Relative-Learning Guarantee}
\label{sec:target-relative-guarantee}

We can now combine the three properties established above to obtain an efficient procedure for estimating all the same-row CDF differences between points in $\Gtar$.

We first construct the mappings $v_j(\cdot)$ using Algorithm~\ref{alg:build-target-maps}.
For each dyadic pair $(a,b)\in\mathcal I^\star$, we consider the dyadic mapped column
\(
    \bigl(v_j(a),v_j(b)\bigr),
\)
and, for each internal index $i$, the correction mapped column
\(
    \bigl(v_j(i),i\bigr).
\)
Each mapped column is partitioned into maximal consecutive blocks over which its endpoints remain constant. On every block having distinct endpoints and at least one eligible row, we apply Algorithm~\ref{alg:bins} with the queryable set restricted to $\Gqry$.
If the endpoints are reversed, we estimate the difference with sorted endpoints and restore its sign. If they coincide, the corresponding difference is zero.
For every pair of target points on the same row, the same-row CDF difference is then reconstructed using the mapped decomposition \eqref{eq:target-mapped-decomposition}.

\begin{proposition}[Row-wise relative learning]
\label{prop:row-relative-learning}
Suppose the input satisfies
\eqref{eq:refine-input-set} and \eqref{eq:refine-input-estimates}, and let
\[
    \varepsilon_{i+1}=\frac{\varepsilon_i}{8}.
\]

There exists a relative-learning procedure that, with probability at least $1-\delta/2$, returns estimates
$\widehat D_j(a,b)$
for every pair $x_{a,j},x_{b,j}\in \Gtar$ sharing the same second coordinate, such that
\[
    \left|
    \widehat D_j(a,b)
    -
    D_j(a,b)
    \right|
    \leq\frac{\varepsilon_{i+1}}{4}.
\]

On the same event, the number of observations is at most
\begin{equation}
\label{eq:row-relative-cost}
    \mathcal O\left(
    \log\frac{24K^3}{\delta}
    \left[
        K\ell_K^2\varepsilon_{i+1}^{-2}
        +
        \ell_K^6\varepsilon_{i+1}^{-3}
    \right]
    \right).
\end{equation}

Every query belongs to $\Gqry$ and has suboptimality gap at most $102\varepsilon_i\ell_K$.
\end{proposition}

\begin{proof}
Set
$\beta=\frac{\varepsilon_{i+1}}{8\ell_K}$ and allocate failure probability $\delta_B=\frac{\delta}{6K^2}$ to each call to Algorithm~\ref{alg:bins}.

There are fewer than $3K^2$ possible calls, independently of the estimation error of the input estimates. Therefore, a union bound guarantees that all elementary CDF differences returned by the binary-search procedures have estimation error at most $\beta$ simultaneously, with probability at least $1-\delta/2$.

By Lemma~\ref{lem:target-availability}, every elementary CDF difference required to reconstruct a same-row comparison between points in $\Gtar$ can be estimated using queryable actions.
Moreover, the mapped decomposition contains at most $2\ell_K$ terms. Hence, the accumulated estimation error is at most
\[
    2\ell_K\beta
    =
    \frac{\varepsilon_{i+1}}{4}.
\]

This proves the precision guarantee.

We now bound the sample complexity.
By \eqref{eq:target-piece-count}, the number of fixed-endpoint blocks is
\[
    |\mathscr P|
    =
    \mathcal O\left(
    K+\frac{\ell_K}{\varepsilon_i}
    \right).
\]

Furthermore, Theorem~\ref{thm:target-overlap} gives
\[
    \mathcal V=\mathcal O(\ell_K^2).
\]

Applying Theorem~\ref{thm:pieces}, with precision $\beta$ and the above failure allocation, gives the bound on the number of one-bit observations
\[
    \mathcal O\left(
    \log\frac{24K^3}{\delta}
    \left[
        \left(K+\frac{\ell_K}{\varepsilon_i}\right)\beta^{-2}
        +
        \ell_K^3\beta^{-3}
    \right]
    \right).
\]

Substituting
\[
    \beta=\frac{\varepsilon_{i+1}}{8\ell_K},
    \qquad
    \varepsilon_i=8\varepsilon_{i+1},
\]
and absorbing lower-order terms yields
\eqref{eq:row-relative-cost}.
Finally, each binary-search call is restricted to $\Gqry$. By Lemma~\ref{lem:threshold-separation}, every queried point has suboptimality gap at most $102\varepsilon_i\ell_K$.
\end{proof}

\section{Constructing the Refinement Oracle}\label{sec:refine}

In this section, we define the internal procedure
$\textsc{Refine}$ and establish the guarantees required
by Definition \ref{def:refinement-oracle}.
Recall that $\textsc{Refine}$ takes as input a set
$\mathcal G\subseteq\mathcal X_K$ and estimates
$\{\widehat F(x)\}_{x\in\mathcal G}$ satisfying
\begin{align*}
\begin{cases}
\displaystyle
\left\{
x\in\mathcal X_K:
\gap(x)\leq c_1\varepsilon_i
\right\}
\subseteq\mathcal G,\\[6pt]
\displaystyle
\max_{x\in\mathcal G}
|\widehat F(x)-F(x)|
\leq\varepsilon_i.
\end{cases}
\end{align*}

The goal is to return a set
$\mathcal \Gtar\subseteq\mathcal X_K$
and corresponding estimates at the improved precision
$\varepsilon_{i+1}=\varepsilon_i/8$, satisfying
\begin{align*}
\begin{cases}
\displaystyle
\left\{
x\in\mathcal X_K:
\gap(x)\leq c_1\varepsilon_{i+1}
\right\}
\subseteq\mathcal \Gtar,\\[6pt]
\displaystyle
\max_{x\in\mathcal \Gtar}
|\widehat F_{\mathrm{new}}(x)-F(x)|
\leq\varepsilon_{i+1}.
\end{cases}
\end{align*}

The procedure proceeds as follows.
First, it uses the input estimates to construct the target set $\Gtar$ and the queryable set $\Gqry$, as defined in \eqref{eq:refine-sets}.
Next, it applies the relative-learning construction developed in the previous sections to estimate the CDF differences between target points on the same row.
Finally, for each row containing at least one target point, it directly estimates the CDF at a single reference point and uses the estimated same-row CDF differences to reconstruct the CDF estimates at all remaining target points.
The procedure returns
$\Gtar$
together with the refined estimates
$\{\widehat F_{\mathrm{new}}(x)\}_{x\in\Gtar}$.

\subsection{Recovering Absolute CDF Estimates}
\label{sec:row-anchors}

The preceding procedure provides all the required CDF differences between target points on the same row. To obtain absolute CDF estimates, it is therefore sufficient to estimate one reference point for each row containing at least one target point.

Let
\[
    \mathcal J_{\mathrm{tar}}
    :=
    \left\{
    j\in\{0,\ldots,K-1\}:
    \exists a,\ x_{a,j}\in \Gtar
    \right\}
\]
be the set of rows containing at least one target point.
For every $j\in\mathcal J_{\mathrm{tar}}$, choose an arbitrary reference point
\[
    a_j=x_{\iota_j,j}\in \Gtar.
\]

We directly estimate $F(a_j)$ using
\[
    n_A=
    \left\lceil
    2\varepsilon_{i+1}^{-2}
    \log\frac{4K}{\delta}
    \right\rceil
\]
fresh one-bit observations.
By Hoeffding's inequality and a union bound over the at most $K$ reference points, with probability at least $1-\delta/2$,
\[
    |\widehat F(a_j)-F(a_j)|
    \leq\frac{\varepsilon_{i+1}}{2},
    \qquad
    \forall j\in\mathcal J_{\mathrm{tar}}.
\]

For every target point $x_{a,j}\in \Gtar$, define
\begin{equation}
\label{eq:row-anchor-reconstruction}
    \widehat F_{\mathrm{new}}(x_{a,j})
    :=
    \operatorname{clip}_{[0,1]}
    \left(
    \widehat F(a_j)
    +
    \widehat D_j(a,\iota_j)
    \right).
\end{equation}

Here $\widehat D_j(a,\iota_j)$ estimates $D_j(a,\iota_j)=F(x_{a,j})-F(a_j)$ and is reconstructed by the relative-learning procedure.
Since clipping cannot increase the estimation error relative to a CDF value in $[0,1]$, the relative-learning and reference-estimation guarantees imply
\begin{align*}
    |\widehat F_{\mathrm{new}}(x_{a,j})
    -F(x_{a,j})|\leq
    |\widehat F(a_j)-F(a_j)|
    +
    \left|
    \widehat D_j(a,\iota_j)
    -
    D_j(a,\iota_j)
    \right|\leq
    \frac{\varepsilon_{i+1}}{2}
    +
    \frac{\varepsilon_{i+1}}{4}
    <\varepsilon_{i+1}.
\end{align*}

Thus, one absolute estimate per nonempty target row is sufficient to reconstruct accurate CDF estimates over the entire target set.
Moreover, since there are at most $K$ such rows, the additional sample complexity is
\[
    \mathcal O\left(
    K\varepsilon_{i+1}^{-2}
    \log\frac{4K}{\delta}
    \right),
\]
which does not change the order of the relative-learning sample complexity.

\subsection{The Complete Row-wise Refinement Procedure}
\label{sec:complete-row-refinement}

We can now combine the mapped relative-learning procedure with the row-wise reference estimates.
The complete procedure is summarized in Algorithm~\ref{alg:row-refine}.
\begin{algorithm}[t]
\small
\caption{\textsc{Refine}$(\mathcal G,\widehat F,\varepsilon_i,\delta)$}
\label{alg:row-refine}
\begin{algorithmic}[1]
\Require Input set $\mathcal G$, estimates $\widehat F$,
precision $\varepsilon_i$, failure probability $\delta$
\Ensure Estimates $\widehat F_{\mathrm{new}}$
on $\Gtar$

\State Set $\varepsilon_{i+1}\gets\varepsilon_i/8$
\State Construct $\Gtar$ and $\Gqry$
       using \eqref{eq:refine-sets}

\State $\{v_j(\cdot)\}\gets
       \Call{BuildTargetMaps}{\Gtar,\Gqry}$

\State Form the dyadic mapped columns and correction mapped columns
\State Partition the mapped columns into maximal fixed-endpoint blocks

\State Set $\beta\gets\varepsilon_{i+1}/(8\ell_K)$
\State Set $\delta_B\gets\delta/(6K^2)$

\For{each fixed-endpoint block $\rho$}
    \State Determine the eligible rows $W_\rho$
    \If{$W_\rho\neq\varnothing$ and the endpoints differ}
        \State Run \Call{BinS}{$\min\{a_\rho,b_\rho\},\max\{a_\rho,b_\rho\},\beta,
               \delta_B,\Gqry,W_\rho$}
        \State Change the sign of the returned estimates if $a_\rho>b_\rho$, then store them in the original orientation
    \EndIf
\EndFor

\For{each $j\in\mathcal J_{\mathrm{tar}}$}
    \State Choose a reference point $a_j=x_{\iota_j,j}\in \Gtar$
           on row $j$
    \State Estimate $F(a_j)$ using $n_A$ fresh observations
\EndFor

\For{each $x_{a,j}\in \Gtar$}
    \State Reconstruct $\widehat D_j(a,\iota_j)$
           using the mapped decomposition
    \State Compute $\widehat F_{\mathrm{new}}(x_{a,j})$
           using \eqref{eq:row-anchor-reconstruction}
\EndFor

\State \Return
$(\Gtar,\{\widehat F_{\mathrm{new}}(x):x\in \Gtar\})$
\end{algorithmic}
\end{algorithm}

We conclude by stating the overall guarantee.

\begin{theorem}[Row-wise refinement]
\label{thm:row-refinement}
Suppose the input satisfies Equations 
\eqref{eq:refine-input-set} and \eqref{eq:refine-input-estimates}, and let
\[
    \varepsilon_{i+1}=\frac{\varepsilon_i}{8}.
\]

With probability at least $1-\delta$,
Algorithm~\ref{alg:row-refine} returns estimates satisfying
\[
    \max_{x\in \Gtar}
    |\widehat F_{\mathrm{new}}(x)-F(x)|
    \leq\varepsilon_{i+1}.
\]

On the same event, its sample complexity is
\begin{equation}
\label{eq:row-refine-sample-complexity}
    \mathcal O\left(
    \log\frac{24K^3}{\delta}
    \left[
        K\ell_K^2\varepsilon_{i+1}^{-2}
        +
        \ell_K^6\varepsilon_{i+1}^{-3}
    \right]
    \right).
\end{equation}

Furthermore, every queried point belongs to $\Gqry$ and has suboptimality gap at most $102\varepsilon_i\ell_K$.

Consequently, on the same event, the regret incurred by the procedure relative to the best grid action is at most
\[
    \mathcal O\left(
    \varepsilon_i\ell_K
    \log\frac{24K^3}{\delta}
    \left[
        K\ell_K^2\varepsilon_{i+1}^{-2}
        +
        \ell_K^6\varepsilon_{i+1}^{-3}
    \right]
    \right).
\]
\end{theorem}

\begin{proof}
By Proposition~\ref{prop:row-relative-learning}, with probability at least $1-\delta/2$, all same-row CDF differences between points in $\Gtar$ are estimated to precision $\varepsilon_{i+1}/4$.

By the reference concentration argument of Section~\ref{sec:row-anchors}, with probability at least $1-\delta/2$, all directly estimated reference values have estimation error at most $\varepsilon_{i+1}/2$.

A union bound shows that both events hold simultaneously with probability at least $1-\delta$.

On this event, the reconstruction
\eqref{eq:row-anchor-reconstruction} gives
\[
    |\widehat F_{\mathrm{new}}(x)-F(x)|
    \leq\frac{3\varepsilon_{i+1}}{4}
    \leq\varepsilon_{i+1}
\]
for every $x\in \Gtar$.

The relative-learning sample complexity follows from
\eqref{eq:row-relative-cost}. The additional $K$ reference estimates require
\[
    \mathcal O\left(
    K\varepsilon_{i+1}^{-2}
    \log\frac{4K}{\delta}
    \right)
\]
observations. This additional cost is absorbed by the first term of
\eqref{eq:row-refine-sample-complexity}.

Finally, all binary-search observations are collected at points in $\Gqry$, and all reference points belong to
$\Gtar\subseteq \Gqry$.

By Lemma~\ref{lem:threshold-separation}, each query has suboptimality gap at most $102\varepsilon_i\ell_K$. Multiplying this bound by the number of observations yields the stated regret guarantee.
\end{proof}

\subsection{Preserving the Refinement Invariant}
\label{sec:repeatable-refinement}

Consider an input $(\mathcal G,\widehat F,\varepsilon_i)$ satisfying
\eqref{eq:refine-input-set} and \eqref{eq:refine-input-estimates}, with
$0<\varepsilon_i\leq1$. The target and queryable sets are defined by
\eqref{eq:refine-sets}. Algorithm~\ref{alg:row-refine} estimates the CDF on
$\Gtar$ to precision $\varepsilon_{i+1}=\varepsilon_i/8$.

The refinement procedure returns $\Gtar$ together with these refined estimates. The following lemma shows that this output already satisfies the input invariant at the new precision.

\begin{lemma}[Preservation of the refinement invariant]
\label{lem:preserve-refinement-invariant}
Suppose that the input satisfies \eqref{eq:refine-input-set} and
\eqref{eq:refine-input-estimates}, and that
\[
    \max_{x\in\Gtar}
    |\widehat F_{\mathrm{new}}(x)-F(x)|
    \leq\varepsilon_{i+1}.
\]
Then
\begin{equation}
\label{eq:new-refinement-invariant}
\begin{aligned}
    &\left\{
    x\in\mathcal{X}_K:
    \gap(x)\leq128\varepsilon_{i+1}\ell_K
    \right\}\subseteq\Gtar,\\
    &\max_{x\in\Gtar}
    |\widehat F_{\mathrm{new}}(x)-F(x)|
    \leq\varepsilon_{i+1}.
\end{aligned}
\end{equation}
\end{lemma}

\begin{proof}
By Lemma~\ref{lem:threshold-separation}, $\Gtar$ contains every grid point with gap at most
$16\varepsilon_i\ell_K=128\varepsilon_{i+1}\ell_K$.
The precision of the refined estimates holds by assumption.
\end{proof}

\begin{comment}
\begin{algorithm}[t]
\small
\caption{\textsc{Refine}$(\mathcal G,\widehat F,\varepsilon_i,\delta)$}
\label{alg:refinement}
\begin{algorithmic}[1]
\Require Nonempty input set $\mathcal G$, estimates $\widehat F$,
precision $\varepsilon_i\in(0,1]$, failure probability $\delta\in(0,1)$
\Ensure Target set $\Gtar$ and refined CDF estimates on $\Gtar$
\State Run $\Call{RowRefine}{\mathcal G,\widehat F,\varepsilon_i,\delta}$
       to obtain $\widehat F_{\mathrm{new}}$ on $\Gtar$
\State \Return $(\Gtar,\widehat F_{\mathrm{new}})$
\end{algorithmic}
\end{algorithm}
\end{comment}

Theorem~\ref{thm:row-refinement} and
Lemma~\ref{lem:preserve-refinement-invariant} immediately give the following guarantee.

\begin{corollary}[Repeatable refinement]
\label{cor:repeatable-refinement}
Suppose that the input to Algorithm~\ref{alg:row-refine} satisfies
\eqref{eq:refine-input-set} and \eqref{eq:refine-input-estimates}.
There is a universal constant $c_0\geq1$ such that, with probability at least $1-\delta$, its output satisfies
\eqref{eq:new-refinement-invariant}, and its number of observations is at most
\begin{equation}
\label{eq:refinement-sample-bound}
    c_0\ell_K^6\log\left(\frac{24K^3}{\delta}\right)
    \left(\varepsilon_{i+1}^{-3}
    +K\varepsilon_{i+1}^{-2}\right).
\end{equation}
Every query belongs to $\Gqry$ and has gap at most
$102\varepsilon_i\ell_K$.
\end{corollary}
Notice that the bound \eqref{eq:refinement-sample-bound} is used only in the analysis; the procedure does not enforce a phase-specific budget. 
Moreover, for any nonempty input set, the target set contains an estimated maximizer. Thus the returned set remains nonempty even when the estimation guarantees fail. 

Corollary~\ref{cor:repeatable-refinement} supplies precisely the refinement oracle required by the successive-elimination analysis in Section~\ref{sec:successive-elimination}, as shown in the following section.

%satisfying Definition~\ref{def:refinement-oracle} with $c_1=128\ell_K=\mathcal{O}(\log(K))$, $c_2=102\ell_K=\mathcal{O}(\log(K))$, and $A(K,\delta)=c_0\ell_K^6\log(24K^3/\delta)=\mathcal{O}(\log^6(K)\log(K/\delta))$.

\section{Proof of Theorem~\ref{thm:main}}

The proof follows directly by combining Corollary~\ref{cor:repeatable-refinement} with
Theorem~\ref{thm:main-parameters}. Indeed, Corollary~\ref{cor:repeatable-refinement} provides a
$(c_1,c_2,A)$-refinement oracle with
\[
c_1=128\ell_K,
\qquad
c_2=102\ell_K,
\qquad
A(K,\delta)=c_0\ell_K^6\log\!\left(\frac{24K^3}{\delta}\right).
\]
With the parameters chosen as in Equation \eqref{eq:complete-parameters}, we have
$\ell_K=\mathcal O(\log T)$ and
$\log(24K^3/\delta)=\mathcal O(\log T)$.
Consequently, $c_2=\mathcal O(\log T)$ and
$A(K,\delta)=\mathcal O(\log^7 T)$.
Applying Theorem~\ref{thm:main-parameters} therefore yields
\[
R_T
=
\mathcal O\!\left(c_2\sqrt{A(K,\delta)}\,T^{2/3}\right)
=
\mathcal O\!\left(T^{2/3}\log^{9/2}T\right),
\]
which proves Theorem~\ref{thm:main}.

\section{Open Problems}\label{sec:discussion}

Our result attains the optimal regret rate for profit maximization in stochastic
bilateral trade up to logarithmic factors. Nonetheless, the following research directions remain open.

\begin{itemize}
    \item \textbf{General CDF-dependent rewards.}
    Our analysis allows an arbitrary joint distribution
    of valuations, but uses the specific profit formula
    $r(x)=(1-x_1-x_2)F(x)$. A natural question is whether
    $\widetilde{\mathcal{O}}(T^{2/3})$ regret can also be
    attained for the broader class of rewards
    $r(x)=g(x)F(x)$, where $g$ is a known bounded
    Lipschitz function, as considered by~\cite{CLM26b}.
    Our bound on the number of samples needed in each
    refinement phase relies on the affine form of the
    profit margin, $1-x_1-x_2$. It is unclear whether
    Lipschitz continuity alone is sufficient to obtain a
    comparable sample bound when this margin is replaced
    by a general function $g$.

    \item \textbf{Adversarial valuations.}
    A second question is whether the same regret rate
    can be attained in the adversarial setting, i.e., when the valuation pairs form an
    arbitrary sequence chosen before the interaction,
    with performance measured against the best fixed
    price pair in hindsight. Our algorithm relies on
    a common distribution across rounds, which allows
    observations collected at different times to
    estimate the same CDF. This also makes it possible
    to eliminate a price permanently once its expected
    profit is sufficiently small. Neither argument
    applies directly when valuations vary
    adversarially. It remains to show whether
    one-bit feedback allows
    $\widetilde{\mathcal{O}}(T^{2/3})$ regret in this
    setting or whether a larger regret
    is unavoidable.
\end{itemize}

\begin{comment}

\paragraph{Time Complexity} Path preparation can be implemented by scanning every junction and row and searching its support for an available replacement. This takes polynomial work in $K$. Every template can then be scanned once per row to form its runs. A comparison is evaluated on demand using its canonical decomposition and stored row lookup; there is no need to precompute all pairwise differences.
%
The graph has at most $K^2$ vertices, and testing its two corners gives an $\mathcal{O}(K^4)$ construction. Greedy selection gives a maximal independent set. These operations, the logarithmic number of phases, and the capped sample count yield an implementation polynomial in $T$.
%
There are two different logarithmic depths. The dyadic support depth controls template incidence, reconstruction length, and overlap across supports. The median-index recursion depth controls repeated probability charges within one \textsc{BinS} call. Both are bounded by $\ell_K$, but they account for different repetitions in the sample bound.
%
The grid used in the theorem has $K-1=2^{\ell_K-1}$ intervals from the outset. An arbitrary desired resolution may be rounded up to the next power of two before running the algorithm, losing at most a factor of two in resolution. No extension to an arbitrary prescribed incomplete grid geometry is needed.
\end{comment}
\newpage
\appendix

%\newpage
%\appendix
\section{Proof of Lemma \ref{lem:grid}}

\lemmaDisc*
\begin{proof}
Fix $x\in[0,1]^2$ with $x_1+x_2\le1$, and let
$u\in\mathcal{X}_K$ be obtained by rounding both coordinates
of $x$ upward. Then $0\le u_i-x_i\le1/(K-1)$ for
$i\in\{1,2\}$ and $F(u)\ge F(x)$.
In terms of prices, this rounding raises the seller's
price and lowers the buyer's price. Trade therefore
becomes at least as likely, while the profit margin
decreases by at most $2/(K-1)$. More precisely,
\begin{align*}
r(x)-r(u)
&=(1-x_1-x_2)\bigl(F(x)-F(u)\bigr)
  +\bigl(u_1-x_1+u_2-x_2\bigr)F(u)\\
&\le \bigl(u_1-x_1+u_2-x_2\bigr)F(u)\\
&\le \frac{2}{K-1},
\end{align*}
where the first inequality uses $x_1+x_2\le1$ and
$F(u)\ge F(x)$, and the second uses $F(u)\le1$.
Since $r(u)\le\OPT$, taking the supremum over $x$
proves the upper bound. The lower bound follows from
$\mathcal{X}_K\subseteq[0,1]^2$, which completes the proof.
\end{proof}

\section{Proof of Lemma \ref{lem:dyadic-decomposition}}

\lemmaPartition*

\begin{proof}
The canonical dyadic decomposition $\mathcal{B}(i,i')$ follows a simple greedy procedure based on the additivity of intervals.

We initialize $\mathcal{B}(i,i')$ as the empty set and start from the smaller endpoint $i'$. At each step, we select the longest dyadic pair $(a,b)\in\mathcal{I}^\star$ such that $a$ coincides with the current endpoint and $b\leq i$. We then add $(a,b)$ to $\mathcal{B}(i,i')$ and continue the procedure from $b$.

Since $\mathcal{I}^\star$ contains all pairs of consecutive indices, there is always an admissible pair as long as the current endpoint is strictly smaller than $i$. The procedure therefore terminates when the right endpoint of the last selected pair coincides with $i$.

By construction, the selected intervals form a consecutive chain from $i'$ to $i$. Consequently, their lengths sum to
\[
    \sum_{(a,b)\in\mathcal{B}(i,i')}(b-a)=i-i'.
\]

It remains to bound the cardinality of $\mathcal{B}(i,i')$.

The greedy construction produces a decomposition into maximal dyadic intervals. A standard property of such a decomposition is that it contains at most two intervals of each dyadic length. %Indeed, the selected intervals correspond to maximal dyadic intervals contained in $(i',i]$, and at each dyadic scale at most two such intervals can occur, one associated with each boundary of the interval.

Since the possible dyadic lengths are $\{1,2,4,\ldots,(K-1)/2\}=\{2^\nu\}_{\nu=0}^{\log_2(K-1)-1}$,
there are $\log_2(K-1)$ possible lengths. Therefore,
\[
    |\mathcal{B}(i,i')|
    \leq 2\log_2(K-1)
    \leq 2\log_2(K),
\]
which concludes the proof.
\end{proof}

\section{Pseudocode and Analysis of the \textsc{BinS} Procedure}
In this section, we provide the pseudocode of \textsc{BinS} and the proof of Lemma \ref{lem:bins}.

\begin{algorithm}[H]
\small
\caption{\textsc{BinS}$(a,b,\beta,\delta_0,\mathcal Q,W)$}
\label{alg:bins}
\begin{algorithmic}[1]
\Require $a<b$ in $\{0,\ldots,K-1\}$, $\beta\in(0,1]$, $\delta_0\in(0,1)$, $\mathcal Q\subseteq\mathcal{X}_K$, $W\neq\varnothing$
\Ensure Estimates $\widehat P_{a,b}(w)$ for all $w\in W$

\State Enumerate the elements of $W$ as $w_0<\cdots<w_M$
\State $n_B\gets\left\lceil32\beta^{-2}\log\frac{4K}{\delta_0}\right\rceil$
\State $\widehat P_{a,b}(w_0)\gets\Call{Estimate}{0}$

\If{$M\ge1$}
    \State $\widehat P_{a,b}(w_M)\gets\Call{Estimate}{M}$
    \State \Call{Split}{$0,M$}
\EndIf

\State \Return $\left\{\widehat P_{a,b}(w):w\in W\right\}$
\par\smallskip\hrule height 0.3pt\smallskip
\Function{Estimate}{$i$}
    \For{$c\in\{a,b\}$}
        \State Post $(z_c,1-z_{w_i})$ for $n_B$ fresh rounds
        \State Set $\widehat F_c(w_i)$ to the fraction of these rounds with trade
    \EndFor
    \State \Return $\widehat F_b(w_i)-\widehat F_a(w_i)$
\EndFunction

\par\smallskip\hrule height 0.3pt\smallskip
\Function{Split}{$i,j$}
    \If{$j=i+1$}
        \State \Return
    \EndIf
    \If{$\widehat P_{a,b}(w_j)-\widehat P_{a,b}(w_i)\le\beta$}
        \State Set
        $\widehat P_{a,b}(w_k)\gets
        \bigl(\widehat P_{a,b}(w_i)+\widehat P_{a,b}(w_j)\bigr)/2$
        for all $i<k<j$
        \State \Return
    \EndIf
    \State $k\gets\lfloor(i+j)/2\rfloor$
    \State $\widehat P_{a,b}(w_k)\gets\Call{Estimate}{k}$
    \State \Call{Split}{$i,k$}
    \State \Call{Split}{$k,j$}
\EndFunction
\end{algorithmic}
\end{algorithm}

\fixend*
\begin{proof}
We first bound the estimation errors at directly sampled rows.

Each invocation of \textsc{Estimate} collects $n_B$ fresh one-bit observations at each endpoint. Conditional on the history preceding each batch, these observations are independent Bernoulli random variables with mean $F(z_c,z_w)$, where $c\in\{a,b\}$ and $w$ is the row being sampled.

By Hoeffding's inequality, the probability that an endpoint estimate deviates from its mean by more than $\beta/8$ is bounded by
\[
    2\exp\left(-2n_B(\beta/8)^2\right)
    \leq\frac{\delta_0}{2K}.
\]
Since each eligible row is directly sampled at most once, the algorithm collects at most $2|W|\leq2K$ endpoint sample means.

Applying a union bound, which remains valid despite the adaptive choice of sampled rows thanks to the conditional concentration bound, we obtain that, with probability at least $1-\delta_0$, all directly sampled endpoint means have estimation error at most $\beta/8$.

Consequently, on this event, subtracting the endpoint estimates gives
\[
    |\widehat P_{a,b}(w)-P_{a,b}(w)|
    \leq\frac{\beta}{4}
\]
simultaneously at all directly sampled rows.

We henceforth condition on this event.

Consider an interval of indices $[i,j]$ on which the algorithm terminates because the stopping condition
\[
    \widehat P_{a,b}(w_j)
    -\widehat P_{a,b}(w_i)
    \leq\beta
\]
is satisfied.

Since both endpoints have estimation error at most $\beta/4$, we obtain
\[
    P_{a,b}(w_j)-P_{a,b}(w_i)
    \leq\beta+2\frac{\beta}{4}
    =\frac{3\beta}{2}.
\]

For any $i<k<j$, the monotonicity of $P_{a,b}$ implies that
\[
    P_{a,b}(w_i)
    \leq P_{a,b}(w_k)
    \leq P_{a,b}(w_j).
\]
Therefore, the distance between $P_{a,b}(w_k)$ and the average of the two true endpoint values is at most half their difference.

Replacing the true endpoint values by their estimates introduces an additional estimation error of at most $\beta/4$. Hence,
\begin{align*}
    &\left|
    \frac{
        \widehat P_{a,b}(w_i)
        +\widehat P_{a,b}(w_j)
    }{2}
    -P_{a,b}(w_k)
    \right|\\
    &\qquad\leq
    \frac12\bigl(P_{a,b}(w_j)-P_{a,b}(w_i)\bigr)
    +\frac{\beta}{4}\\
    &\qquad\leq
    \frac{3\beta}{4}+\frac{\beta}{4}
    =\beta.
\end{align*}

Thus, all estimates assigned without direct sampling have estimation error at most $\beta$. Directly sampled rows already have estimation error at most $\beta/4$, while intervals containing no interior eligible rows require no additional estimates. This proves the estimation guarantee simultaneously for every $w\in W$.

It remains to bound the number of directly sampled rows.

Whenever the algorithm subdivides an interval $[i,j]$, its estimated increment must satisfy
\[
    \widehat P_{a,b}(w_j)
    -\widehat P_{a,b}(w_i)>\beta.
\]
Using the estimation error bounds at the two endpoints, the corresponding true increment satisfies
\[
    P_{a,b}(w_j)-P_{a,b}(w_i)
    >\beta-2\frac{\beta}{4}
    =\frac{\beta}{2}.
\]

By the definition of $P_{a,b}$, this increment is precisely
\[
    P_{a,b}(w_j)-P_{a,b}(w_i)
    =
    \mathbb{P}\bigl(
    X\in(z_a,z_b]\times(z_{w_i},z_{w_j}]
    \bigr).
\]

At any fixed recursion depth, the half-open intervals $(w_i,w_j]$ associated with distinct recursive calls are disjoint and contained in $(\min W,\max W]$. Consequently, the sum of their probability masses is at most
\[
\mathbb{P}\bigl(X\in(z_a,z_b]\times(z_{\min W},z_{\max W}]\bigr).
\]

Since each interval that is subdivided has probability mass greater than $\beta/2$, the number of subdivisions at any fixed recursion depth is at most
\[
    \frac{2\mathbb{P}\bigl(X\in(z_a,z_b]\times(z_{\min W},z_{\max W}]\bigr)}{\beta}.
\]

Moreover, each subdivision splits the set of eligible rows approximately in half. Since $|W|\leq K$, the recursion has at most $\lceil\log_2 K\rceil$ levels.

Each subdivision introduces exactly one new directly sampled row. Accounting for the at most two initially sampled rows, we conclude that the total number of directly sampled rows is bounded by
\[
    2+
    \frac{2\lceil\log_2 K\rceil}{\beta}
    \mathbb{P}\bigl(X\in(z_a,z_b]\times(z_{\min W},z_{\max W}]\bigr),
\]
which proves \eqref{eq:bins-evaluations} and concludes the proof.
\end{proof}

\newpage

\section*{AI methodology}
We used generative AI tools (ChatGPT 6, OpenAI) to generate an initial version of the proof. We subsequently revised and refined the arguments, reconciling them with our own understanding of the problem and the partial proof we had independently developed. We also used generative AI tools to polish the writing for clarity and readability. The authors take full responsibility for the correctness of the final results and proofs.

\printbibliography[title={References}]
\end{document}